\documentclass[aps,pra,twocolumn,showpacs,superscriptaddress]{revtex4-1}   
\usepackage{graphicx}  
\usepackage{epstopdf}
\usepackage[caption=false]{subfig}
\usepackage{float} 
\usepackage[toc,page]{appendix}
\usepackage{cleveref}
\usepackage{dcolumn}   
\usepackage{bm}        
\usepackage{amssymb}   
\usepackage{mathrsfs}   
\usepackage{color}
\usepackage{hyperref}
\usepackage{mathtools}
\usepackage[table]{xcolor}
\usepackage{amsmath, amsthm, amssymb}
\usepackage{amsfonts}
\usepackage{bbold}       
\usepackage{accents}    

\hypersetup{
    colorlinks=true, 
    linktoc=all,     
    linkcolor=blue,  
    citecolor=blue,
    filecolor=blue,
    urlcolor=blue
}

\usepackage{soul}
\usepackage{cancel}

\newcommand{\beq}{\begin{equation}}
\newcommand{\eeq}{\end{equation}}
\newcommand{\bqa}{\begin{eqnarray}}
\newcommand{\eqa}{\end{eqnarray}}
\newcommand{\bse}{\begin{subequations}}
\newcommand{\ese}{\end{subequations}}
\newcommand{\nn}{\nonumber}

\newcommand{\erf}[1]{Eq.~(\ref{#1})}

\newcommand{\erfa}[2]{Eqs.~(\ref{#1}) and (\ref{#2})}
\newcommand{\arf}[1]{Appendix~\ref{#1}} 
\newcommand{\trf}[1]{Table~\ref{#1}} 
\newcommand{\srf}[1]{Sec.~\ref{#1}}

\newcommand{\frf}[1]{Fig.~\ref{#1}}

\newcommand{\ie}{{\it i.e.~}}

\newcommand{\dg}{^\dagger}
\newcommand{\expt}[1]{\langle{#1}\rangle}

\definecolor{BLACK}{gray}{0}
\definecolor{dgray}{rgb}{0.26,0.26,0.26}
\definecolor{RED}{rgb}{1,0,0}
\definecolor{GREEN}{rgb}{0.2,.5,0.2}
\definecolor{BLUE}{rgb}{0,0,1}
\definecolor{AWESOME}{rgb}{0.54,0.77,0.79}
\definecolor{BEN}{rgb}{0.44,0.37,0}

\renewcommand{\(}{\left(}
\renewcommand{\)}{\right)}
\newcommand{\sq}[1]{\left[{#1}\right]}

\newcommand{\abs}[1]{\left| {#1} \right|}
\newcommand{\tr}[1]{{\rm Tr}\sq{ {#1} }}

\newcommand{\smallfrac}[2]{\mbox{$\frac{#1}{#2}$}}

\newcommand{\bra}[1]{\langle{#1}|}
\newcommand{\ket}[1]{|{#1}\rangle}
 
\newcommand{\op}[2]{\ket{#1}\bra{#2}}
\newcommand{\s}[1]{ \hat{\sigma}_{#1}}

\newcommand{\gc}{ {\rm g} }
\newcommand{\er}{ {\rm e} }

 \newcommand{\gu}[1]{\gamma_{\uparrow,#1}}
 \newcommand{\gd}[1]{\gamma_{\downarrow, #1}}

   \newcommand{\gdu}{\gamma_{\downarrow\uparrow}}
   \newcommand{\gdd}{\gamma_{\downarrow\downarrow}}

\newcommand{\opvec}[1]{\hat{\mathbf{#1}}}
\newcommand{\opmat}[1]{\hat{\mathbf{#1}}}
\newcommand{\mat}[1]{\mathbf{#1}}

\begin{document}

\widetext

\

\title{Quantum master equations for entangled qubit environments}

\author{Shakib Daryanoosh} \email{sh.daryanoosh@gmail.com}
\affiliation{Centre for Engineered Quantum Systems, Department of Physics and Astronomy, \\ Macquarie University, Sydney, NSW 2122, Australia}
\author{Ben Q. Baragiola} 
\affiliation{Centre for Quantum Computation and Communication Technology, School of Science, \\ RMIT University, Melbourne, Victoria 3001, Australia}
\affiliation{Centre for Engineered Quantum Systems, Department of Physics and Astronomy, \\ Macquarie University, Sydney, NSW 2122, Australia}
\author{Thomas Guff}
\affiliation{Centre for Engineered Quantum Systems, Department of Physics and Astronomy, \\ Macquarie University, Sydney, NSW 2122, Australia}
\author{Alexei Gilchrist} 
\affiliation{Centre for Engineered Quantum Systems, Department of Physics and Astronomy, \\ Macquarie University, Sydney, NSW 2122, Australia}

%
%


%
\vskip 0.25cm

\date{\today}

\begin{abstract} 
We study the Markovian dynamics of a collection of $n$ quantum systems coupled to an irreversible environmental channel consisting of a stream of $n$ entangled qubits. 
Within the framework of repeated quantum interactions, we derive the master equation that describes the dynamics of the composite quantum system. 
We investigate the evolution of the joint system for two-qubit environments and find that (1) the presence of antidiagonal coherences (in the local basis) in the environment is a necessary condition for entangling two remote systems, and (2) that maximally entangled two-qubit baths are an exceptional point without a unique steady state. 
For the general case of $n$-qubit environments we show that coherences in maximally entangled baths (when expressed in the local energy basis), do not affect the system evolution in the weak coupling regime.


\end{abstract}

\maketitle

\section{Introduction}

Open quantum systems are the subject of extensive research since physical quantum systems cannot be entirely isolated from their surroundings. The influence of the environment often manifests as unwanted noise that can thwart attempts to exploit intrinsic quantum properties for quantum computing, communication, and metrology \cite{BarKim17}.  
Open systems tend to lose their key quantum properties---coherence and entanglement---as they interact with the environment. This is not inevitable, however, and much research has focused on \emph{engineered} environments \cite{MulLin12} for various tasks including quantum computing \cite{VerCir09,PasCir11,LegDev15} and the generation of novel steady states \cite{LinWin13, KieHom14, BarTwa18}. This can be achieved through a combination of precise structuring of system-bath interactions and  preparation of the environment in particular states. 

 One can take many approaches to the description of open quantum systems. In standard quantum optical treatments, the electromagnetic field serves as the environment \cite{Car08, WisMil10, BrePet03}, and the dynamics of the reduced quantum states is given by a master equation (ME) after the environmental degrees of freedom are traced out. 
An alternative approach is that of repeated quantum interactions \cite{Brun:2002aa, Pel10b, BruMer14}, also called collision models \cite{Rau:1963aa, Scarani:2002aa,GioPal12, LorPal17, Francesco:2017aa}, which treats the system-environment coupling discretely. 
The environment is comprised of a chain of 
identical and independent quantum ancillae which sequentially couple to the system and are then traced out.
Taking a continuous limit \cite{AttJoy07}, the resulting dynamical map on the reduced system state becomes a ME. 
This formalism has been applied in the context of quantum thermodynamics \cite{Hor12, LorPal15, DagKur16, StrEsp17, LiSha18}, as well as quantum optics and information \cite{GroCom17}. 
The framework of repeated quantum interactions has also proven useful for the study of correlated quantum channels \cite{Korotkov:2002aa, GioPal12,TomasRybar:2012aa}. When the ancillae are entangled, 
the reduced-state dynamics can exhibit non-Markovian behavior \cite{Kretschmer:2016aa, LorPal17}, such as propagating single-excitation states \cite{DabChr17}.

\begin{figure}
\includegraphics[scale=0.5]{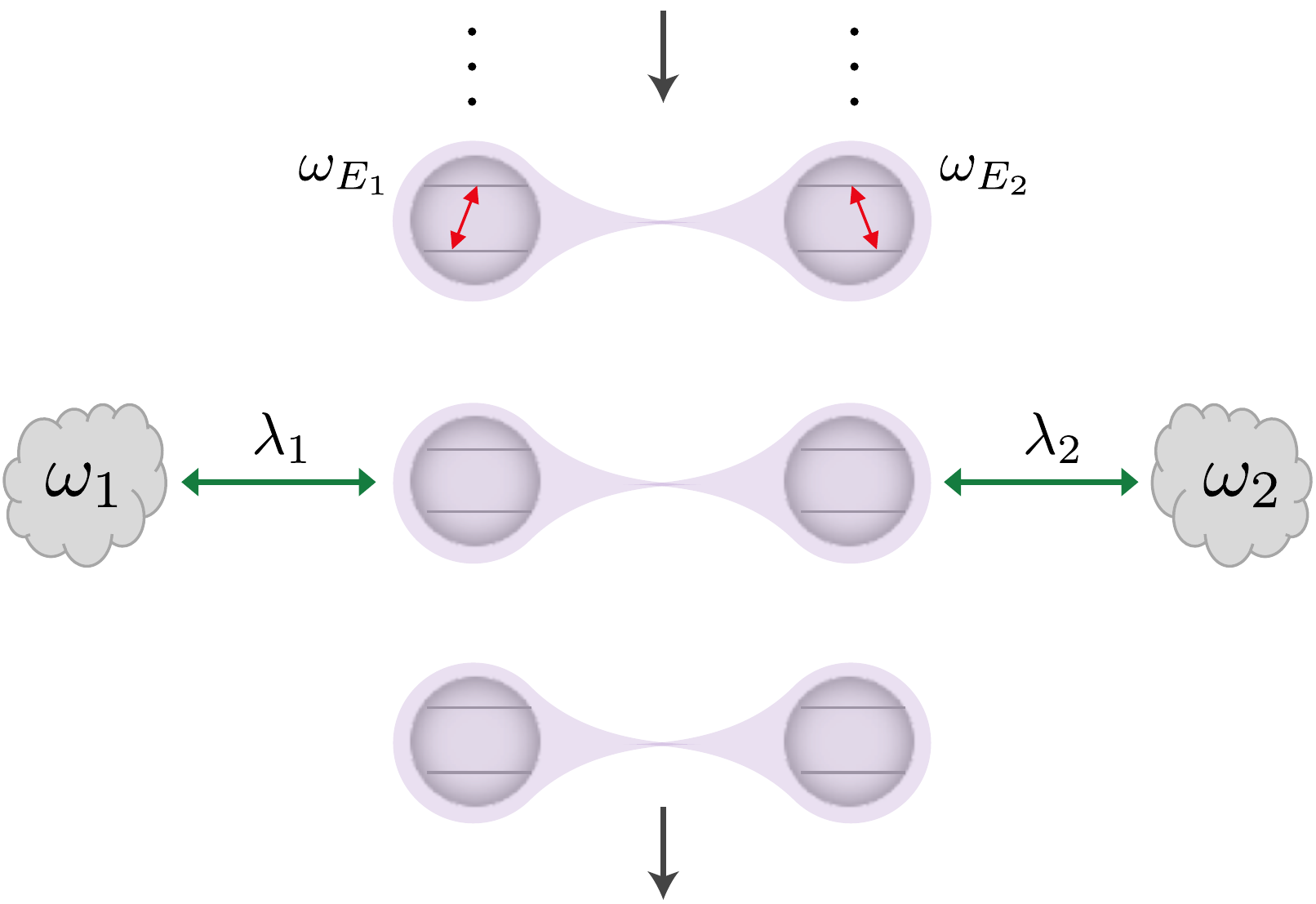}
\caption{\label{fig:setup} (Color online). Conceptual diagram of the physical model where a stream of entangled qubits sequentially interacts with separate quantum systems. Shown here is the case of $n=2$ qubits. }
\end{figure}

In this article we consider a correlated environment that interacts with many quantum systems simultaneously. The environment consists of a stream of $n$ ancilla qubits, each coupled to its own system. The $n$ qubits arrive entangled with one another at each interaction time, but they are not entangled across different times, which allows us to derive a Markovian ME for the joint state of the quantum systems. Depending on the state of the qubits, the ME can generate non-trivial dynamics of the reduced systems.  
This provides the tools to tackle the problem of converting coherences and/or entanglement in the environment 
into quantum correlations in the system \cite{WuGu17, CerSou17, RieGro18}. 
Accordingly, we analyze in detail the pedagogical case of $n=2$ qubits in the bath, which can be prepared as a stream of entangled states, as it provides the canonical method for transferring  qubit entanglement to system entanglement. When the bath is prepared arbitrarily close to a maximally entangled state, the system is driven to an entangled pure state, such as a two-mode squeezed state for the case of two optical cavities.  Surprisingly, if the qubit bath is \emph{exactly} in a Bell state, the system fails to converge to a unique steady state. 
For $n$-qubit baths $(n>2)$, we find that 
 for certain multi-qubit entangled environments such as baths prepared in $X$-states (which have nonzero elements only on the diagonal and antidiagonal entries of the state matrix when expressed in the local energy basis) \cite{Wei10, RafEbe12}, the bath entanglement \emph{cannot} be transferred to the systems, as a direct consequence of the weak-coupling limit

The manuscript is organized as follows.
In \srf{sec:physmod} the underlying framework behind our analysis is explained.
We first present and interpret the two-qubit bath ME in \srf{sec:ME:2q}. We give two forms for the ME, which are useful for pure-state and general mixed-state baths, respectively. 
In \srf{sec:2qb} and \srf{sec:2cav} we apply the formalism to study the dynamics and steady states for two sets of quantum system: optical cavities and two-level atoms. 
The general case of an $n$-qubit bath is dealt with in \srf{sec:ME}. Finally, in \srf{sec:con} we summarize the findings of this paper and propose future directions.

\section{Repeated interaction with a bath of $n$ entangled qubits} \label{sec:physmod}
We derive dynamical maps and master equations within the structure of repeated quantum interactions \cite{Brun:2002aa, BruMer14, StrEsp17, GroCom17}. In this formalism, a quantum system in Hilbert space ${\mathscr H}_{S}$ couples to an environment which is comprised of a stream of identical and independent quantum systems such that ${\mathscr H}_{E} \equiv \bigotimes_ {l} {\mathscr H}_{E}^{(l)}$. We assume the environment has infinitely many elements, although in principle it can be finite. Each environmental element sequentially interacts with the system over a short time interval of duration $\Delta t = t_l - t_{l+1}$ after which it no longer interacts with the system. 
Tracing over the environmental degrees of freedom 
yields a map on the system of interest in ${\mathscr H}_{S}$.
This is similar to the standard scenario for open systems in quantum optics where a bosonic probe field interacts with the system in a continuous-in-time manner \cite{WisMil10}. However, the situation here is different in two ways. 
First, the system-environment coupling is fundamentally discrete, although we will ultimately consider short-time interactions and take a continuous limit \cite{AttPau06, AttJoy07, Pel08}. 
Second, the environmental systems are qubits rather than bosonic modes. This serves not only to model physical situations where streams of qubits interact with a fixed quantum system \cite{NogHar99}, but in addition the results fit into the framework of quantum computing and simulation \cite{GroCom17}.

For each interaction time interval $\Delta t$ the total Hilbert space of the system plus the segment of the environment interacting at that time is ${\mathscr H} = {\mathscr H}_{S} \otimes  {\mathscr H}_{E}^{(l)}$, and the Hamiltonian is
\beq
	\hat{H} = \hat H_0 + \hat{H}^{(l)}_I,
\eeq
corresponding to the bare Hamiltonians of the system and of the environment, $\hat{H}_0$, and the the system-environment interaction, $\hat{H}_I^{(l)}$. 
We take the environment to be comprised of a stream of entangled qubits with each qubit coupled to 
its own quantum system, see \frf{fig:setup}. That is, in each time interval $\Delta t$, $n$ entangled qubits interact with $n$ quantum systems, which themselves are left arbitrary and can in general be remote from each other.   Note that the bath qubits within a single time interval are entangled, but they are not entangled \emph{between} time intervals. This type of environmental entanglement drives non-Markovian dynamics \cite{Korotkov:2002aa, Gough:2011aa, Baragiola:2012aa, DabChr17, Baragiola:2017aa} and will be treated separately.

Each subsystem interacts with its respective qubit via a coupling operator $\hat{c}_j$. The bare and interaction Hamiltonians in the rotating wave approximation are 
\bqa
	\hat H_0 &=& \sum_{\ell=1}^n \(  \omega_{\ell} \hat c_{\ell}\dg \hat c_{\ell} + \omega_{\textit{\tiny E}_\ell} \s{\ell}\dg \s{\ell}\), \\
	\hat{H}^{(l)}_I &=& \sum_{\ell=1}^n \lambda_\ell \( \hat c_\ell \s{\ell}\dg + \hat c_\ell\dg \s{\ell}\),
\eqa
where $\omega_{\ell}$ and $\omega_{\textit{\tiny E}_\ell}$ are the respective transition frequencies of the subsystems and the bath qubits, $\lambda_\ell$ is the coupling strength between the $\ell$-{th} subsystem and its qubit, and the bath qubit lowering operator is $\s{\ell} = \ket{\gc}_\ell\bra{\er}$. 
In the interaction picture with respect to the bare Hamiltonian, $\hat H_0$, 
the joint unitary evolution is generated by the time-dependent Hamiltonian,
\beq  \label{IPIH:2q}
	\hat{H}^{(l)}_I(\Delta t) = \sum_{\ell=1}^n \lambda_\ell \( \hat c_\ell\dg \s{\ell} e^{-i \delta_\ell \Delta t} + {\rm H. c.}\),
\eeq
where $\delta_\ell \coloneqq \omega_{\textit{\tiny E}_\ell} - \omega_{\ell}$ is the detuning. 
The detuning is included here for completeness; henceforth, we focus our attention to resonant system-qubit interactions, $\delta_\ell = 0$. From the interaction, one models a specific reservoir by selecting a particular state for the environmental qubits. Investigating situations where the bath qubits are entangled is the focus of this article.

A dynamical map for the joint state of the $n$ fixed quantum systems is found by tracing out the the environmental qubits after the interaction, $\hat{U}_I^{(l)}$, generated by the Hamiltonian in \erf{IPIH:2q}. At each time interval the incoming bath qubits and the quantum systems are assumed to be in a product state, so the dynamical map is completely-positive and trace-preserving. Assuming that only one qubit interacts with each system in an interaction time and the joint state of the $n$-qubit environment  interacting in the time interval is $\hat \rho_E$, the dynamical map is given by
\beq \label{rhoS:gen}
	\hat \rho{(t_{l+1})} = {\rm Tr}_E \sq{ \hat{U}_I^{(l)} \big( \hat \rho{(t_l)} \otimes \hat \rho_E^{({l})} \big) \hat{U}_I^{(l)}{}\dg}.
\eeq
The fact that each set of $n$ qubits, described by $\hat{\rho}_E$, is independent from other sets means that this dynamical map is Markovian (\emph{i.e.} arises from a memoryless environment). This is because the joint state of the input environment is a tensor product state across interaction time intervals, $\hat{\Psi}_{E} \equiv \bigotimes_ {l} {\hat{\rho}}_{E}^{(l)}$ \cite{Baragiola:2017aa}. Thus, the dynamical map at every subsequent time interval is of the form of \erf{rhoS:gen}, using the system state from the previous time interval and a fresh environmental state $\hat{\rho}_E$.
From now on, for the ease of notation we drop the explicit superscripts for the system and environment state unless confusion could arise.

\section{Master equations for two-qubit baths} \label{sec:ME:2q}

In this section we focus on two-qubit baths as the quintessential extension of the single-qubit baths that are typically studied \cite{LorPal15, StrEsp17, GroCom17, DabChr17, LiSha18}. Two-qubit baths can exhibit nonclassical correlations including maximal entanglement. We investigate how two-qubit baths modify correlations between two remote subsystems---optical cavities in \srf{sec:2cav} and two-level atoms in \srf{sec:2qb}. Rather than using the discrete dynamical maps in \erf{rhoS:gen}, we take a continuous-time limit and describe the reduced-state dynamics by a Markovian ME. 

A ME can be derived from \erf{rhoS:gen} under a set of standard assumptions. 
First, each environmental qubit spends the same amount of time $\Delta t$ interacting with its local system. Second, the rotating wave approximation, which has already been made in \erf{IPIH:2q}, requires that the interaction time is long in comparison to the system's characteristic time $\omega_{\ell} \Delta t \gg 1$. Third, the Markov condition requires that the $n$-qubit environment in each time interval is independent of other intervals. Finally, the system-bath coupling is weak $\lambda_\ell < \omega_{\ell}$ while $\lambda_\ell \Delta t \ll 1$. 
In the weak-coupling regime the unitary time-evolution operator is expanded up to second order in $\Delta t$,
\bse
\bqa 
	\hat{U}_I^{(l)} &=& e^{-i \hat{H}^{(l)}_I \Delta t} \\ 
&=& \hat{\mathbb 1}_{SE} - i \hat{H}^{(l)}_I \Delta t - \frac{1}{2} (\hat{H}^{(l)}_I)^2 {\Delta t}^2 + {\mathit O}({\Delta t}^3). \label{Utaylor}
\eqa
\ese
Let us assume that in each time interval the two bath qubits are prepared in the pure state $\hat \rho_E = \op{\psi_E}{\psi_E}$, where
\beq \label{rhoE:pure}
\ket{\psi_E} = b_{\er\er} \ket{\er\er} + b_{\er\gc} \ket{\er\gc} + b_{\gc \er} \ket{{\gc}\er} + b_{\gc \gc} \ket{\gc \gc},
\eeq
and the coefficients satisfy
\beq
b_{jk} \in {\mathbb C}, \quad {\rm and} \quad \sum_{j,k \in \{\er,\gc\}} \abs{b_{jk}}^2 =1.
\eeq

We insert \erf{Utaylor} into the dynamical map for the reduced system state, \erf{rhoS:gen}, and then evaluate the terms under the two-qubit bath state in \erf{rhoE:pure}. 
The ME in \erf{ME:gen}, which describes the continuous limit of repeated quantum interactions, then arises in the limit of infinitesimal interaction, $\Delta t \rightarrow dt$. 
This calculation, whose details can be found in Appendix \ref{appnA1}, yields the following master equation ($\hbar = 1)$, for the reduced state $\hat{\rho}$:
\begin{align} \label{ME:gen}
       \dot{\hat \rho}(t) 
        & =- i [ \hat{H}^{\rm eff}_1 + \hat{H}^{\rm eff}_2, \hat \rho] + \sum_{m=1}^4 \mathcal{D}[ \hat L_m] \hat \rho
\end{align}
where the effective Hamiltonians are
\begin{subequations} \label{Hams:gen}
   \bqa
         \hat{H}^{\rm eff}_1 &=& \lambda_1 (b_{\gc\gc} b_{\er\gc}^\ast + b_{\gc\er} b_{\er\er}^\ast) \hat{c}_1 + {\rm H. c.}, \\
         \hat{H}^{\rm eff}_2 &=& \lambda_2 (b_{\er\gc} b_{\er\er}^\ast+ b_{\gc\gc} b_{\gc\er}^\ast ) \hat{c}_2 + {\rm H. c.},
    \eqa
\end{subequations}
the jump operators are
	\begin{subequations}  \label{eq:jumpops}
	\begin{align}
		\hat L_1 =& \sqrt{ \gamma_1 } b_{\gc \gc} \hat{c}_1 + \sqrt{ \gamma_2 } b_{\er \er} \hat{c}_2\dg , \\
		\hat L_2 =& \sqrt{ \gamma_1 } b_{\er \er} \hat{c}_1\dg + \sqrt{ \gamma_2 } b_{\gc \gc} \hat{c}_2 , \\
		\hat L_3 =& \sqrt{ \gamma_1 } b_{\gc \er} \hat{c}_1 + \sqrt{ \gamma_2 } b_{\er \gc} \hat{c}_2 , \\
		\hat L_4 =& \sqrt{ \gamma_1 } b_{\er \gc} \hat{c}_1\dg + \sqrt{ \gamma_2 } b_{\gc \er} \hat{c}_2\dg , 
	\end{align}
	\end{subequations}
and the Lindblad superoperator is defined as
	\beq
		{\cal D}[\hat o] \hat{\rho} = \hat o \hat{\rho} \hat o\dg - \smallfrac{1}{2} \{ \hat o\dg \hat o , \hat{\rho}\}_+, 
	\eeq
with $\{\hat{A}, \hat{B} \}_+ = \hat{A} \hat{B} + \hat{B} \hat{A}$. 
The relative rates are given by $ \gamma_\ell= |\lambda_\ell|^2 \Delta t$.

The master equation generates both coherent dynamics and incoherent dynamics in the reduced system state $\hat{\rho}(t)$. The Hamiltonian terms in \erf{Hams:gen} arise from coherences within each separate bath qubit. The coherent, unitary dynamics they generate are analogous to coherent driving \cite{WisMil10}. Simultaneously, the quantum system undergoes correlated dissipation as described by the jump operators in \erf{eq:jumpops}. Each jump operator drives a dissipative process given by combinations of loss ($\hat c_\ell$) and heating ($\hat c\dg_\ell$) across subsystems 1 and 2. 

Interestingly, the jump operators~\eqref{eq:jumpops} are determined by the state in two two-dimensional subspaces of the qubit bath, spanned by either $\{\ket{\gc \gc}, \ket{\er \er} \}$ or $\{\ket{\gc \er}, \ket{\er \gc} \}$. Equivalently, each subspace is spanned by two Bell states. Populations and coherences within subspace $\{\ket{\gc \gc}, \ket{\er \er} \}$ contribute to jump operators $\hat L_1$ and $\hat L_2$, and similarly populations and coherences within subspace $\{\ket{\gc \er}, \ket{\er \gc} \}$ contribute to jump operators $\hat L_3$ and $\hat L_4$. Coupling between the subspaces is due to certain single-qubit coherences that manifest in the effective Hamiltonians in \erf{Hams:gen}.

\subsection{Unentangled bath qubits} \label{prod:bath}

An elementary situation is the case of unentangled bath qubits, $\ket{\psi_E}=\ket{\psi_1} \otimes \ket{\psi_2}$. Consider uncorrelated bath qubits, both in the ground state, $b_{\gc \gc} =1$. The result is the typical ME for independent loss on each subsystem through the jump operators $\hat L_1 = \sqrt{\gamma_1} \hat{c}_1$ and $\hat L_2 = \sqrt{\gamma_2} \hat{c}_2$, with $\hat L_3 = \hat L_4 = 0$. The situation becomes more complicated if either bath qubit contains excitation. In this case the ME can involve all four jump operators in \erf{eq:jumpops} even when the two bath qubits are uncorrelated. 
Consider the case where each qubit is in the state $\ket{\psi_\ell} = \alpha_\ell \ket{\gc} + \beta_\ell \ket{\er}$ with $|\alpha_\ell|^2 + |\beta_\ell|^2=1$, and with no initial subsystem correlations, $\hat{\rho}= \hat{\rho}_1 \otimes \hat{\rho}_2$. The ME becomes
\begin{align} \label{}
       \dot{\hat{\rho}}(t) 
         = & -i  \big[ \hat{H}^{\rm eff}_1, \hat{\rho}_1]\otimes \hat{\rho}_2 - i\, \hat{\rho}_1 \otimes [ \hat{H}^{\rm eff}_2, \hat{\rho}_2 ] \\
        & + \sum_{\ell=1,2}\gamma_\ell \big( |\alpha_\ell|^2 \mathcal{D}[\hat{c}_\ell] +|\beta_\ell|^2 \mathcal{D}[\hat{c}_\ell\dg] \big) \hat{\rho}_1 \otimes \hat{\rho}_2 \nn \\
                & - \sqrt{\gamma_1 \gamma_2} [ \hat{H}^{\rm eff}_1, \hat{\rho}_1 ] \otimes [ \hat{H}^{\rm eff}_2, \hat{\rho}_2] \nn ,
\end{align}
with effective Hamiltonians (for $\ell=1,2$),
   \bqa \label{ham:eff:prod}
         \hat{H}^{\rm eff}_{\ell} &=&  \lambda_\ell \alpha_\ell \beta_\ell^* \hat{c}_\ell + {\rm H. c.} 
    \eqa
The first two lines describes the independent evolution of each subsystem, as would be expected if one were to separately derive and add together the ME for each subsystem. 
Here the ME contains an additional term in the final line. It describes a classically correlated coherent driving of the two subsystems, but it does not generate entanglement and preserves the separable structure of the joint state, $\hat{\rho}(t) = \hat{\rho}_1(t) \otimes \hat{\rho}_2(t)$. Further, tracing over either subsystem gives the expected single-subsystem ME.  Note that this term does not appear in typical bosonic MEs due to vanishing bath correlation functions.

 \subsection{Bath qubits in a general state} 
While the master equation in \erf{rhoE:pure} was derived for a two-qubit bath in a pure state, it is also valid for the environment in a general two-qubit state,
\beq \label{eq:twoqubitstate}
\hat{\rho}_E = \sum_{j,k} \sum_{j', k'} b_{jk,j'k'} \op{jk}{j'k'},
\eeq
with indices $j,k, j', k' \in \{\er,\gc \}$. In order to show the correspondence we transform the ME into another useful form by expanding \erf{ME:gen} and collecting terms according to coefficients, we get 
  \begin{align} \label{ME:gen:2q}
       \dot{\hat{\rho}}(t) &= - i [ \hat{H}^{\rm eff}_1 + \hat{H}^{\rm eff}_2, \hat{\rho}]  \\
        & + \sum_{\ell=1}^2 \gd{\ell}  {\cal D}[\hat{c}_\ell] \hat{\rho} + \sum_{\ell=1}^2 \gu{\ell}  {\cal D}[\hat{c}_\ell\dg] \hat{\rho} \nonumber \\
        & +\, \gdd \,{\cal S}[\hat{c}_1,\hat{c}_2] \hat{\rho} + \gdd^* \,{\cal S}[\hat{c}_1\dg,\hat{c}_2\dg] \hat{\rho}     \nonumber \\
        & + \gdu \,{\cal S}[\hat{c}_1,\hat{c}_2\dg] \hat{\rho} +\, \gdu^* \,{\cal S}[\hat{c}_1\dg,\hat{c}_2] \hat{\rho}. \nn 
\end{align}
We have defined a superoperator, 
	\begin{align}
		\mathcal{S}[\hat{o}_1,\hat{o}_2]\hat{\rho} \coloneqq \hat{o}_1 \hat{\rho} \hat{o}_2 + \hat{o}_2 \hat{\rho} \hat{o}_1 - \smallfrac{1}{2} \{ \hat{o}_1 \hat{o}_2 + \hat{o}_2 \hat{o}_1, \hat{\rho} \}_+,
	\end{align}
that is symmetric in the arguments.
The coefficients in \erf{ME:gen:2q} are given by
\begin{subequations} \label{gammas:gen}
   \bqa
         \gd{1} &=& \gamma_1 \big( \abs{b_{\gc\gc}}^2 + \abs{b_{\gc\er}}^2 \big), \label{gd1}  \\
         \gu{1} &=& \gamma_1  \big(  \abs{b_{\er\er}}^2 + \abs{b_{\er\gc}}^2 \big),  \label {gu1}  \\
         \gd{2} &=& \gamma_2 \big(  \abs{b_{\gc\gc}}^2 + \abs{b_{\er\gc}}^2 \big), \label {gd2} \\
         \gu{2} &=& \gamma_2 \big(  \abs{b_{\er\er}}^2 + \abs{b_{\gc\er}}^2 \big), \label {gu2} \\
         \gdd &=& \sqrt{ \gamma_1 \gamma_2}  \, b_{\gc\gc} b_{\er\er}^\ast, \label{gdd} \\
         \gdu &=&  \sqrt{ \gamma_1 \gamma_2} \, b_{\gc \er} b_{\er\gc}^\ast. \label{gdu}
    \eqa
\end{subequations}
The first four coefficients in Eqs.~(\ref{gd1}-\ref{gu2}) are positive and can be interpreted as rates.  The final two \erfa{gdd}{gdu} are complex and the superoperator terms that they multiply may in general interfere with other terms including the local dissipators. This is indeed a consequence of \erf{ME:gen:2q} not being in diagonal form with respect to the jump operators. For a general state of the two-qubit bath,  \erf{eq:twoqubitstate}, we replace the coefficients in \erf{gammas:gen} according to 
\[
b_{jl}\, b_{j'l'}^\ast \rightarrow b_{jl,j'l'},
\]
and use these coefficients in the ME given in \erf{ME:gen:2q}.
The MEs in \erf{ME:gen} and \erf{ME:gen:2q} are identical, but each may be more useful for certain calculations.

\begin{figure}
\centering
\includegraphics[scale=0.37]{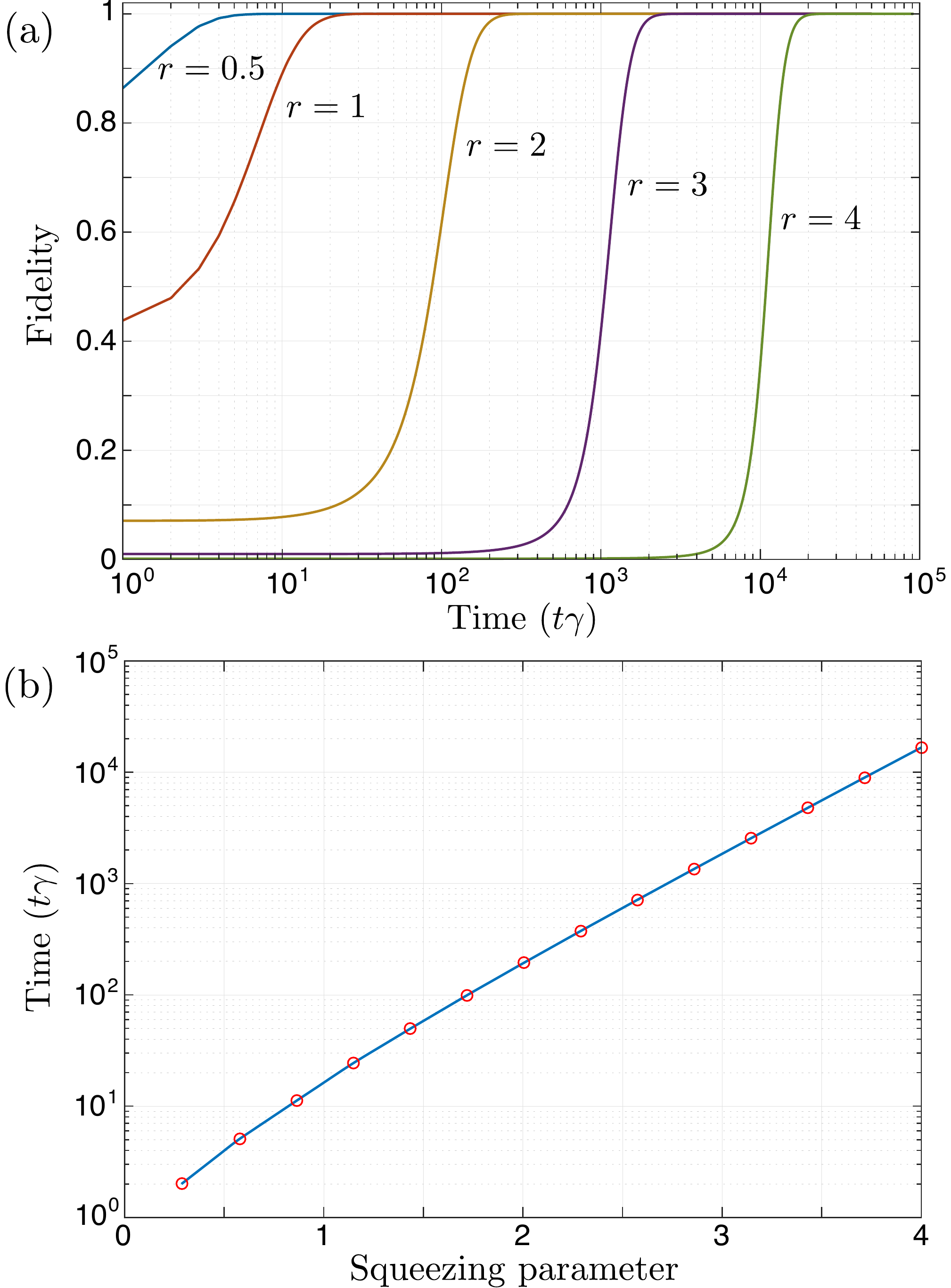}
\caption{\label{fig:squeezing} (Color online) Dissipative preparation of a two-mode squeezed vacuum state across two remote cavities. The joint cavity state, initially prepared in two-mode vacuum, evolves under the ME in \erf{ME:gen} with jump operators given by \erf{2mode:op}. The associated squeezing parameter $r$ is related to the bath qubit coefficients, see \erf{eq:squeezingparam}. For the plotted values of $r$, the ground state coefficient magnitudes are $|b_{\gc \gc}| \approx \{ 0.908, 0.796, 0.720, 0.709, 0.707\}$ and the squeezing angle is set to $\vartheta = 0$. Smaller $|b_{\gc \gc}|$ corresponds to larger squeezing and infinite squeezing to $|b_{\gc \gc}| = 1/\sqrt{2}$.  (a) Fidelity of the state with the two-mode squeezed vacuum state \erf{eq:TMSV} for various values of $r$. (b) Time at which the fidelity surpasses 0.98 as a function of the squeezing parameter $r$. }
\end{figure}

\section{Two remote cavities: two-mode squeezing} \label{sec:2cav} 

A stream of qubits interacting with a harmonic oscillator is the prototype for a variety of tasks. For example, two-level Rydberg atoms interacting with an ultra-high-finesse microwave cavity, have been used for quantum nondemolition measurements of photon number \cite{NogHar99} and stabilization of Fock states in the cavity \cite{Sayrin:2011aa}. We consider here the natural extension of this system to two remote cavities 
\cite{CamDev17,DidMir17,RieGro18}. Each subsystem is a single-mode cavity with mode operators $\hat{a}_\ell$ and $\hat{a}_m\dg$ that satisfy canonical commutation relations, $[\hat{a}_\ell, \hat{a}_m\dg] = \delta_{\ell m}$.  
We now consider an arbitrary two-qubit bath state in the two-dimensional subspace spanned by $\{\ket{\gc \gc}, \ket{\er \er}\}$,
\begin{align} \label{phi:gen}
\ket{\psi_E} = b_{\gc\gc} \ket{\gc\gc}  + b_{\er\er} \ket{\er\er} .
\end{align}
That is, the coefficients in \erf{rhoE:pure} take values $b_{\er\gc} = b_{\gc\er} = 0$, while $|b_{\gc\gc}|^2 + |b_{\er \er}|^2 = 1$. For simplicity each cavity couples to its respective stream of bath qubits with the same rate ($\lambda_1 = \lambda_2$) via an interaction that exchanges excitations, which corresponds to $\hat{c}_\ell \rightarrow \hat{a}_\ell$. 
Thus the ME in \erf{ME:gen} has $\hat{L}_3 = \hat{L}_4 = 0$ and 
\begin{subequations}  \label{2mode:op}
\bqa
\hat{L}_1 & = & \sqrt{\gamma} \(b_{\gc \gc} \hat{a}_1 +  b_{\er \er} \hat{a}_2\dg\), \label{sq:op1} \\
\hat{L}_2 & = & \sqrt{\gamma} \( b_{\gc \gc} \hat{a}_2 +  b_{\er \er} \hat{a}_1\dg \), \label{sq:op2}
\eqa
\end{subequations}
reminiscent of two-mode squeezing transformation.
 
The connection to two-mode squeezing can be made explicit under certain conditions. When $|b_{\gc \gc}| > 1/\sqrt{2}$ 
we can define a strictly positive effective rate given by the population difference,
\beq \label{eq:effectiverate}
\Gamma \coloneqq  \gamma \(|b_{\gc \gc}|^2 - |b_{\er \er}|^2 \).
\eeq
Then, the jump operators can be rewritten as 
\begin{subequations} \label{sq:op}
\bqa 
\hat{L}_1 & = & \sqrt{\Gamma} \Big[ {\rm cosh}(r) \hat{a}_1 + e^{i \vartheta }\, {\rm sinh}(r) \hat{a}\dg_2 \Big] , \label{sq:op1} \\
\hat{L}_2 & = & \sqrt{\Gamma} \Big[ {\rm cosh}(r) \hat{a}_2 + e^{i \vartheta }\, {\rm sinh}(r) \hat{a}\dg_1 \Big] , \label{sq:op2}
\eqa
\end{subequations}
where the squeezing amplitude $r$ is related to the coefficients via the relations, 
\beq \label{eq:squeezingparam}
\cosh (r)  = \frac{|b_{\gc \gc}|}{\sqrt{ |b_{\gc\gc}|^2 - |b_{\er\er}|^2}}.  
\eeq
 The squeezing angle $\vartheta$ is given by the phase of $b_{\er \er}$ relative to $b_{\gc \gc}$. Preparing the qubit bath in a maximally entangled state, $|b_{\gc \gc}| = 1/2$, is the limit of infinite squeezing, $r \rightarrow \infty$. 
The jump operators are explicitly given by a two-mode squeezing transformation on the cavity annihilation operators, $\hat{L}_\ell  =  \sqrt{\Gamma} \hat{S} \hat{a}_\ell \hat{S}\dg$, where the unitary, two-mode squeezing operator is \cite{Wang:2007aa, Sch17}
\beq \label{eq:squeezeop}
	\hat{S} 
		= e^{\zeta^* \hat{a}_1 \hat{a}_2 - \zeta \hat{a}_1\dg \hat{a}_2\dg},       
\eeq
with complex squeezing parameter $ \zeta = r e^{i \vartheta}$. The transformation is in the Schr\"{o}dinger-picture because the jump operators are \emph{nullifiers} \cite{Gu:2009aa,Menicucci:2011aa} of the two-mode squeezed vacuum state. This can be seen directly by transforming the action of the annihilation operators,
	\begin{align}
		\hat{a}_\ell \ket{0}_2 = 0 \rightarrow \hat{S}\hat{a}_\ell \hat{S}\dg \hat{S} \ket{0}_2 \propto \hat{L}_\ell \ket{r,\vartheta}_2 = 0,
	\end{align}
where the two-mode squeezed vacuum state is
\beq \label{eq:TMSV}
	\ket{r, \vartheta}_2 \coloneqq \hat{S} \ket{0}_2 .
\eeq
Thus, the steady state of the ME is a pure, Gaussian, two-mode squeezed vacuum state $\ket{r, \vartheta}_2$ with squeezing that depends on the qubit-bath coefficients. 
The dynamical preparation of $\ket{r, \vartheta}_2$ from a two-mode vacuum state is shown for various values of the squeezing parameter $r$ in \frf{fig:squeezing}(a). The state is dissipatively cooled via interaction with the two-qubit bath towards the steady state, details can be found in \arf{appendix:Gaussian}. We quantify the approach to $\ket{r, \vartheta}_2$ with the Uhlmann-Jozsa fidelity, which can be calculated from the covariance matrix. As $r$ is increased, the effective rate $\Gamma$ decreases according to \erf{eq:effectiverate}. That the time to approach the steady state $\ket{r, \vartheta}_2$ increases exponentially with the squeezing parameter $r$---see \frf{fig:squeezing}(b)---is unsurprising, since more highly squeezed states contain more energy. 

In the opposite regime, where $|b_{\gc \gc}| < 1/\sqrt{2}$, there is no unitary Bogoliubov transformation that transforms the operators $\hat{a}_j$ while maintaining the canonical commutation relations. 
Nevertheless, the jump operators in \erf{2mode:op} may be written similarly to \erf{sq:op} with the roles of $\cosh (r)$ and $\sinh (r)$ reversed. 
Because $|\cosh (r)/\sinh (r) | > 1$, the jump operators contain a larger proportion of $\hat{a}_\ell\dg$ than $\hat{a}_\ell$, and the incoming two-qubit environment is more likely to transfer energy to the subsystems than to remove it. In this case the ME serves as an incoherent amplifier. In the following section we will investigate this parameter regime as well as the ``exceptional points" where the bath is prepared in a maximally entangled state such as a Bell state. 

Note that for the case with coefficients $b_{\gc\gc} = b_{\er\er} = 0$ does not lead to the same dynamics even though the amount of entanglement in the bath can be the same. In this case $\hat{L}_1 = \hat{L}_2 = 0$ and the resulting master equation just has dissipative terms not related to squeezing.

\section{Two remote two-level atoms interacting with a Bell-state bath} \label{sec:2qb}

In this section we investigate the repeated interaction between two remote two-level subsystems and a stream of maximally entangled bath qubits. The subsystems are taken to be identical, each described by a bare Hamiltonian $\hat{H}_{\ell} = \frac{\omega_0}{2} \hat{\sigma}_{z,\ell}$, where $\omega_0$ is resonant with the bath qubit frequency, $\delta_\ell = \omega_{E_\ell} - \omega_0 = 0$. The interaction between each subsystem and its bath qubit is an excitation exchange described by a lowering operator, $\hat{c}_\ell \rightarrow \s{\ell}$. To avoid confusion, we henceforth refer to the bath as \emph{qubits} and the subsystems as \emph{atoms}, with the understanding that the bath qubits could indeed be physically manifested as a stream of entangled atoms. 

\subsection{Bath qubits in a pure Bell state} \label{Sec:Bellbath}

We first consider a maximally entangled two-qubit bath state in the subspace $\{\ket{\gc \gc}, \ket{\er \er} \}$,
\beq \label{phi:plusminus}
\ket{\psi_E} = \frac{1}{\sqrt{2}} \Big( \ket{\er\er} + e^{i\phi} \ket{\gc\gc} \Big),
\eeq
such that the Bell states $\ket{\Phi_E^+}$ and $\ket{\Phi_E^-}$ are given by $\phi = (0,\pi)$, respectively.
That is, the coefficients in \erf{rhoE:pure} take values $b_{\er \er} = 1/\sqrt{2}$ and $b_{\gc\gc} = e^{i \phi}/\sqrt{2}$ while $b_{\er\gc} = b_{\gc\er} = 0$. For simplicity we set $\lambda_1 = \lambda_2 = \lambda$ corresponding to decay rate $\gamma$, which yields the ME:
\begin{align}  \label{ME:Bell:phi}
	\dot{\hat{\rho}}(t) =&  \frac{\gamma}{2} \left( \mathcal{D} \Bigl[ \s{1} + e^{i\phi } \s{2}\dg \Bigr]\hat{\rho} +  \mathcal{D} \Bigl[ e^{i\phi}\s{1}\dg + \s{2} \Bigr]\hat{\rho} \right) .
\end{align}
The joint state of the atomic subsystems is initialized in the arbitrary state
\beq
	\hat{\rho}_0 = \sum_{j,k} \sum_{j',k'} \rho_{jk,j'k'}^0 \op{jk}{j'k'},
\eeq
where the sums run over $\{\gc, \er\}$.

\subsubsection{Bath qubits in $\ket{\Phi_E^+}$}

First, we consider the case where the bath qubits are prepared in the Bell state, $\ket{\Phi_E^+}$, given by $\phi = 0$ in \erf{phi:plusminus}.
In the long-time limit, $t \rightarrow \infty$, the steady state of the two-atom system is given by
 \beq  \label{rhoSS:phiplus}
\hat{\rho}_{\rm ss} =  \left( 
\begin{array}{cccc}
\rho_{\er\er,\er\er}^{\rm ss}  & 0 & 0 & \rho_{\er\er,\gc\gc}^{\rm ss}  \\ 
0 & \rho_{\er\gc,\er\gc}^{\rm ss} & 0 & 0 \\
0 & 0 & \rho_{\gc\er,\gc\er}^{\rm ss} & 0 \\
\rho_{\er\er,\gc\gc}^{{\rm ss}^\ast} & 0 & 0 &  \rho_{\gc\gc,\gc\gc}^{\rm ss} 
\end{array}
\right),
\eeq
where the steady-state matrix elements are related to the initial-state matrix elements by
\begin{eqnarray}
\rho_{\er\er,\er\er}^{\rm ss} &=& \smallfrac{1}{3} \big[ \rho_{\er\er,\er\er}^0 + \rho_{\gc\gc,\gc\gc}^0 - \rho_{\er\er,\gc\gc}^0 + \smallfrac{1}{2} \rho_{\er\gc,\er\gc}^0 + \smallfrac{1}{2} \rho_{\gc\er,\gc\er}^0 \big], \nonumber \\
\rho_{\er\gc,\er\gc}^{\rm ss} &=& \smallfrac{1}{3} \big[ \smallfrac{1}{2}\rho_{\er\er,\er\er}^0 + \smallfrac{1}{2} \rho_{\gc\gc,\gc\gc}^0 + \rho_{\er\er,\gc\gc}^0 + \rho_{\er\gc,\er\gc}^0 + \rho_{\gc\er,\gc\er}^0 \big], \nonumber \\
\rho_{\gc\er,\gc\er}^{\rm ss}  &=& \rho_{\er\gc,\er\gc}^{\rm ss}, \nonumber\\
\rho_{\gc\gc,\gc\gc}^{\rm ss} &=&\rho_{\er\er,\er\er}^{\rm ss}, \nonumber \\
\rho_{\er\er,\gc\gc}^{\rm ss} &=& -\smallfrac{1}{6} \big[\rho_{\er\er,\er\er}^0 + \rho_{\gc\gc,\gc\gc}^0 - 4 \rho_{\er\er,\gc\gc}^0 - \rho_{\er\gc,\er\gc}^0 - \rho_{\gc\er,\gc\er}^0 \big]. \nonumber
\end{eqnarray}
Thus the atomic steady state is not unique; rather, the ME has an invariant subspace. A particular steady-state within this invariant subspace depends on the initial state \cite{Albert:2014aa}. 

The steady state of the joint atomic system can exhibit non-classical correlations, identified by the negative partial transpose criterion \cite{Per96}. The partial transpose matrix $\rho_{\rm ss}^{\rm PT}$, partitioned with respect to the subsystems, takes the following form,
 \beq  \label{rhoSS:PT}
\hat{\rho}_{\rm ss}^{\rm PT} =  \left( 
\begin{array}{cccc}
\rho_{\er\er,\er\er}^{\rm ss}  & 0 & 0 & 0  \\ 
0 & \rho_{\er\gc,\er\gc}^{\rm ss} &  \rho_{\er\er,\gc\gc}^{\rm ss}  & 0 \\
0 & \rho_{\gc\gc,\er\er}^{\rm ss}  & \rho_{\gc\er,\gc\er}^{\rm ss} & 0 \\
0& 0 & 0 &  \rho_{\gc\gc,\gc\gc}^{\rm ss} 
\end{array}
\right).
\eeq
Negativity in the spectrum of $\hat{\rho}_{\rm ss}^{\rm PT}$ guarantees the presence of entanglement. We quantify the entanglement by the logarithmic negativity \cite{Ple05}, 
\beq \label{log:neg}
	LN(\hat{\rho}) := {\rm log}_2 \(\tr{\sqrt{\hat{\rho}_{\rm PT}\dg \hat{\rho}_{\rm PT}}}\),
\eeq
where $LN(\hat{\rho}) \in [0,1]$ with the minimum value corresponding to separable states and the maximum value to maximally entangled states.

A particular steady state of interest is when the atomic system is prepared in the Bell state $\ket{\Phi^-}$, whence it does not undergo any decoherence via interaction with the qubit bath, since it is already at the steady state. However, when the initial atomic state is $\ket{\Phi^+}$ the steady state is the mixture, 
\beq \label{rhoss:phipp}
	\hat{\rho}_{\rm ss} = \smallfrac{1}{3}\(\op{\Phi^+}{\Phi^+} + \op{\er\gc}{\er\gc} +\op{\gc\er}{\gc\er}\),
\eeq
with a positive partial transpose matrix, \ie it has purely positive eigenvalues $(\frac{1}{2}, \mathbf{\frac{1}{6}})$ with boldface indicating degeneracy of order 3. A positive partial transpose is a sufficient condition for separability of a two-qubit system. 

An interesting scenario is an initial joint atomic state, $\hat{\rho}_0(\theta) = \op{\psi_0(\theta)}{\psi_0(\theta)}$, that is a weighted sum of the two Bell states,
\beq  \label{phi:plus:minus}
\ket{\psi_0(\theta)} = \sin \theta \ket{\Phi^+} + \cos \theta \ket{\Phi^-},
\eeq
for $0 \le \theta \le \pi/2$.  Beyond this range in $\theta$, the entanglement behavior repeats. Above we found that the maximally entangled state, $\hat \rho_0(\theta = 0) = \op{\Phi^-}{\Phi^-}$, is a steady state of the ME
As $\theta$ deviates from zero, the contribution from the antisymmetric Bell state diminishes. Beyond the critical point $\theta_c = \pi/4$ ($\ket{\psi_0} = \ket{\er \er}$) the effect of the symmetric Bell state is dominant and the entanglement vanishes, $LN[\hat{\rho}_{\rm ss}(\theta\ge\theta_c)] = 0$. At the critical point the steady state is a two-atom Werner state \cite{Wer89},
\beq \label{rho:werner}
\hat{\rho}_{\rm ss}(\theta_c) 
	= \frac{1}{3} \op{\Phi^-}{\Phi^-} + \frac{1}{6} \hat{\mathbb 1}_S,
\eeq
which is separable. In fact, even after this point the steady state is separable all the way to $\theta=\pi/2$, at which point it is given by \erf{rhoss:phipp}. The behavior is illustrated in \frf{fig:log_neg}(a) where we plot logarithmic negativity as a function of $\theta$ (dashed blue curve). Comparison with the initial logarithmic negativity (light gray curve) reveals that the Bell state environment preserves the initial system entanglement up until $\theta_c$ despite the fact that the state becomes mixed, \frf{fig:log_neg}(b). 

\subsubsection{Bath qubits in $\ket{\Phi_E^-}$, $\ket{\Psi_E^+}$, or $\ket{\Psi_E^-}$} 
The situation where the bath is prepared in $\ket{\Phi_E^-}$ proceeds similarly. In this case, the steady state has the same form as \erf{rhoSS:phiplus} with the following substitutions:
\beq
\rho_{\er\er,\gc\gc}^{0} \rightarrow - \rho_{\er\er,\gc\gc}^{0}, \quad {\rm and} \quad \rho_{\er\er,\gc\gc}^{\rm ss} \rightarrow - \rho_{\er\er,\gc\gc}^{\rm ss}.
\eeq
When the initial atomic state is parameterized as in \erf{phi:plus:minus}, the steady states for the extremal cases, $\theta = (0,\theta_c)$, are just as in \erf{rhoss:phipp} and \erf{rho:werner}, respectively, with the roles of $\ket{\Phi^{\pm}}$ swapped.
By varying $\theta$ in the interval $[0,\theta_c]$, the antisymmetric Bell state is dominant, the result of which is that the systems remain separable. 
For the range $\theta > \theta_c$ the symmetric Bell state, $\ket{\Phi^+}$ is dominant in the initial atomic state, and the ME preserves the initial entanglement. For $\theta = \pi/2$ the initial state, $\rho_0(\theta = \pi/2) = \op{\Phi^+}{\Phi^+}$, is a steady state.

Lastly, when the qubit environment is prepared in the other subspace $\{\ket{\er\gc}, \ket{\gc\er}\}$,
\beq \label{psi:plusminus}
\ket{\psi_E} = \frac{1}{\sqrt{2}} \Big( \ket{\er\gc} + e^{i\phi} \ket{\gc\er} \Big),
\eeq 
with Bell states $\ket{\Psi_E^+}$ and $\ket{\Psi_E^-}$ given by $\phi = \{0,\pi\}$, the steady states have an analogous form. \trf{tbl:psi:pm} summarizes the results for comparison.

\begin{figure}
\centering
\includegraphics[scale=0.47]{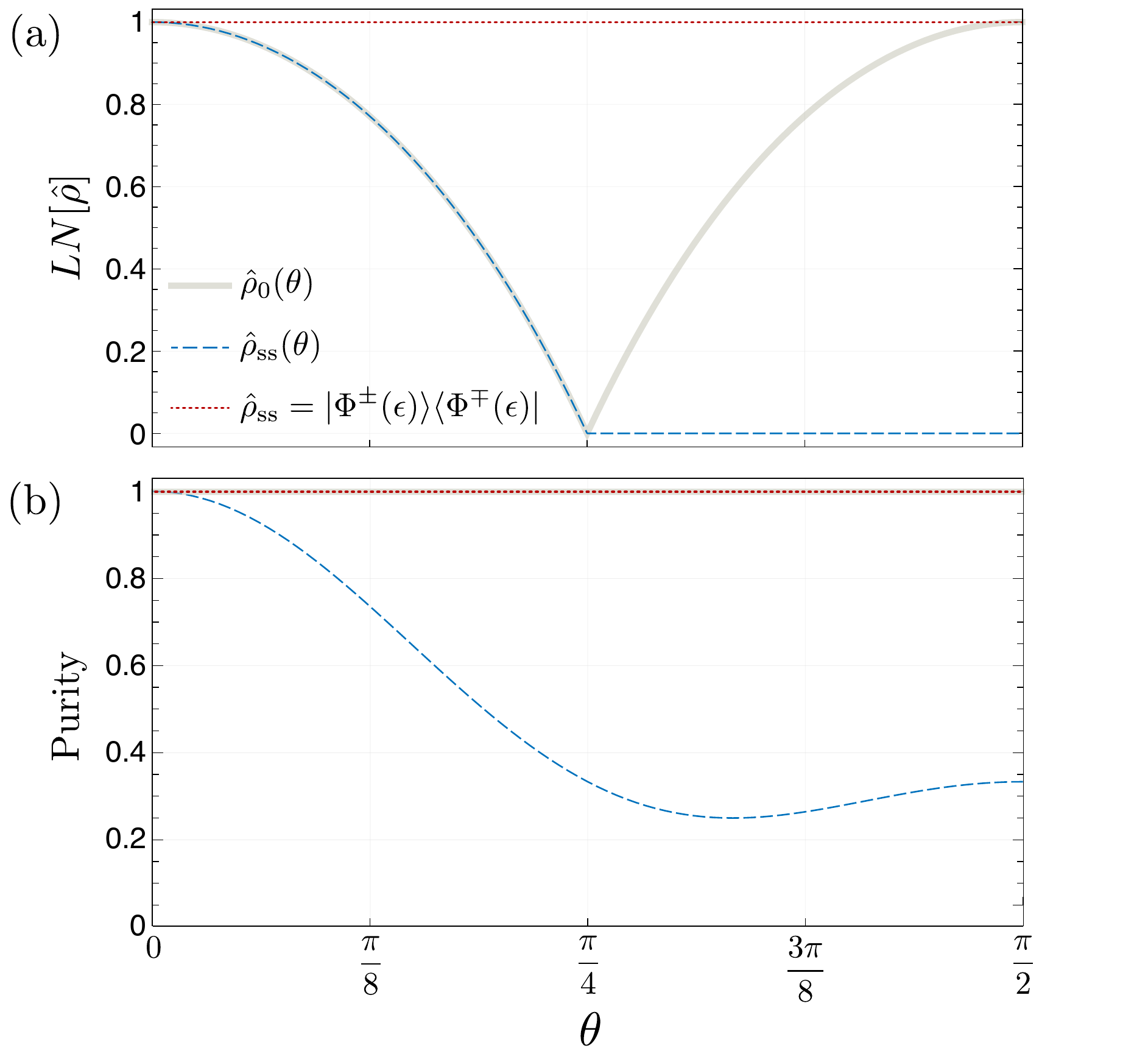}
\caption{\label{fig:log_neg} (Color online). Entanglement and purity of the atomic state when the two-qubit bath is prepared in the $\ket{\Phi^+_E}$ Bell state. The initial two-atom state $\hat{\rho}_0(\theta)$ is parameterized by $\theta$, \erf{phi:plus:minus}. We also include plots for non-maximally entangled atomic states, $\ket{\Phi^\pm(\epsilon)}$, that deviate from Bell states by $\epsilon = 0.001$, \erf{eq:nearBell}.  (a) Entanglement as quantified by the logarithmic negativity, \erf{log:neg}, for which $LN(\hat{\rho}) \approx 1$. (b) State purity, $\tr{\hat{\rho}^2}$. Note that for $\theta<\theta_c = \pi/4$, the gray curve and dashed blue curve exactly coincide. That is, initial entanglement is preserved despite the fact that the state becomes mixed. Beyond $\theta_c$ the steady state is no longer entangled, and the minimum purity, 0.25, occurs at $\theta = \pi/3$.  
}
\end{figure}

\begin{table*}
	\caption{\label{tbl:psi:pm} Summary of results when the environment is prepared in a Bell state or a state very close to a Bell state, $\ket{\Phi^{\pm}(\epsilon)}$, \erf{eq:nearBell}, or $\ket{\Psi^\pm(\epsilon)} \equiv \big(\ket{\er\gc} \pm \sqrt{1+\epsilon}\, \ket{\gc\er} \big)/\sqrt{2+\epsilon}$ parametrized by $0<\epsilon \ll 1$. The initial atomic state, \erf{phi:plus:minus}, is parameterized by $\theta$, and numbers in bold indicate three-fold degeneracy in the eigenvalue. When the two-qubit environment is \emph{exactly} prepared in either of the two Bell states, the steady state $\hat{\rho}_{\rm ss}$ depends on the initial state $\hat{\rho}_0$ and the atoms remain entangled only for some specific parameter regimes. However, for an environment initialized in $\ket{\Phi^{\pm}(\epsilon)}$ or $\ket{\Psi^{\pm}(\epsilon)}$, the two atoms can acquire nonclassical correlations arbitrarily close to maximally entangled states. 
 }
	\begin{tabular}{ c | c  c  c  c}
		\hline \hline 
		 \begin{tabular}{@{}c@{}} {\rm Two-qubit bath state}\;\; \\ $\hat{\rho}_E$ \end{tabular}  & \;\;$\theta$\;\; &  \begin{tabular}{@{}c@{}} {\rm Initial  atomic state} \\ $\hat{\rho}_0(\theta)$ \end{tabular} & \begin{tabular}{@{}c@{}} {\rm Steady state} \\$\hat{\rho}_{\rm ss}(\theta)$ \end{tabular}    & \begin{tabular}{@{}c@{}} {\rm Spectrum of} $\hat{\rho}_{\rm ss}^{\rm PT}(\theta)$ \end{tabular} \\  \hline
		 $\op{\Phi^+}{\Phi^+}$  & 0 & $\op{\Phi^-}{\Phi^-}$ & $\op{\Phi^-}{\Phi^-}$ & $\frac{1}{2} (-1,{\mathbf 1})$   \\[4pt] 
		" & $\theta_c= \frac{\pi}{4}$  & $\op{\er\er}{\er\er}$  & $\frac{1}{6} \hat{\mathbb 1}_S + \frac{1}{3} \op{\Phi^-}{\Phi^-}$  & $ (\frac{1}{2},\mathbf{\frac{1}{6}})$ \\ [4pt] 
		" & $\frac{\pi}{2}$ & $\op{\Phi^+}{\Phi^+}$ & $\frac{1}{3}\(\op{\er\gc}{\er\gc} + \op{\gc\er}{\gc\er} + \op{\Phi^+}{\Phi^+}\)$ & $ \frac{1}{3}	 (0, \mathbf{1})$  \\ [4pt] \hline 
		$\op{\Phi^-}{\Phi^-}$ & 0 & $\op{\Phi^-}{\Phi^-}$ & $\frac{1}{3}\(\op{\er\gc}{\er\gc} + \op{\gc\er}{\gc\er} + \op{\Phi^-}{\Phi^-}\)$ & $\frac{1}{3} (0, \mathbf{1})$ \\ [4pt] 
		 " &$\theta_c$ & $\op{\er\er}{\er\er}$ & $\frac{1}{6} \hat{\mathbb 1}_S + \frac{1}{3} \op{\Phi^+}{\Phi^+}$ & $(\frac{1}{2}, \mathbf{\frac{1}{6}})$ \\ [4pt] 
		 " & $\frac{\theta}{2}$ & $\op{\Phi^+}{\Phi^+}$ & $\op{\Phi^+}{\Phi^+}$ & $\frac{1}{2} (-1,{\mathbf 1})$  \\[4pt] \hline 
		  $\op{\Phi^\pm(\epsilon)}{\Phi^\pm(\epsilon)}$ & $\in [0,\frac{\pi}{2}]$ & $\op{\psi_0(\theta)}{\psi_0(\theta)}$ & $\op{\Phi^\mp(\epsilon)}{\Phi^\mp(\epsilon)}$ & $(\frac{1}{2+\epsilon}, \frac{1+\epsilon}{2+\epsilon}, \frac{1}{2}, -\frac{1}{2})$ \\ [4pt]
		\hline 
		$\op{\Psi^+}{\Psi^+}$  & 0 & $\op{\Psi^-}{\Psi^-}$ & $\op{\Psi^-}{\Psi^-}$ & $\frac{1}{2} (-1,{\mathbf 1})$   \\[4pt] 
		" & $\theta_c$  & $\op{\er\gc}{\er\gc}$  & $\frac{1}{6} \hat{\mathbb 1}_S + \frac{1}{3} \op{\Psi^-}{\Psi^-}$  & $ (\frac{1}{2},\mathbf{\frac{1}{6}})$ \\ [4pt] 
		" & $\frac{\pi}{2}$ & $\op{\Psi^+}{\Psi^+}$ & $\frac{1}{3}\(\op{\er\er}{\er\er} + \op{\gc\gc}{\gc\gc} + \op{\Psi^+}{\Psi^+}\)$ & $ \frac{1}{3}	 (0, \mathbf{1})$  \\ [4pt] \hline 
		$\op{\Psi^-}{\Psi^-}$ & 0 & $\op{\Psi^-}{\Psi^-}$ & $\frac{1}{3}\(\op{\er\er}{\er\er} + \op{\gc\gc}{\gc\gc} + \op{\Psi^-}{\Psi^-}\)$ & $\frac{1}{3} (0, \mathbf{1})$ \\ [4pt] 
		 " &$\theta_c$ & $\op{\er\gc}{\er\gc}$ & $\frac{1}{6} \hat{\mathbb 1}_S + \frac{1}{3} \op{\Psi^+}{\Psi^+}$ & $(\frac{1}{2}, \mathbf{\frac{1}{6}})$ \\ [4pt] 
		 " & $\frac{\theta}{2}$ & $\op{\Psi^+}{\Psi^+}$ & $\op{\Psi^+}{\Psi^+}$ & $\frac{1}{2} (-1,{\mathbf 1})$  \\[4pt] \hline 
		  $\op{\Psi^\pm(\epsilon)}{\Psi^\pm(\epsilon)}$ & $\in [0,\frac{\pi}{2}]$ & $\op{\psi_0(\theta)}{\psi_0(\theta)}$ & $\op{\Psi^\mp(\epsilon)}{\Psi^\mp(\epsilon)}$ & $(\frac{1}{2+\epsilon}, \frac{1+\epsilon}{2+\epsilon}, \frac{1}{2}, -\frac{1}{2})$ \\ [4pt]
		\hline \hline
	\end{tabular}
\end{table*}

\subsection{Bath qubits in a non-maximally entangled state}

From the above analysis it may be inferred that the existence of coherences in the environment is merely a necessary condition for entangling the subsystems, even though the cross terms in the master equation, \erf{ME:Bell:phi}, might suggest otherwise, \ie the generation of quantum correlation among systems.  We found that Bell-state baths generate a ME without a unique steady state and, depending on the initial atomic state, atomic entanglement is either preserved or destroyed, but never created. Here we show that when the bath qubits are prepared in a non-maximally entangled state that can be arbitrarily close to a Bell state, the ME has a unique, entangled steady state. Consider the bath in the following state that slightly deviates from a maximally entangled state, 
\beq \label{phi:plus:eps}
\ket{\psi_E(\phi,\epsilon)} = \frac{1}{\sqrt{2+\epsilon}} \Big(\ket{\er\er} + e^{i\phi}\sqrt{1+\epsilon}\, \ket{\gc\gc} \Big).
\eeq
 
The atomic steady state is highly entangled for all $\phi$, with logarithmic negativity, $LN(\hat{\rho}) \approx 1$ for $0 < \epsilon \ll 1$.
This can be seen by considering the following \emph{approximate} Bell states for the qubit bath:
	\begin{subequations} \label{eq:nearBell}
	\begin{align}	
		\ket{\Phi^{+}(\epsilon)} \coloneqq & \ket{\psi_E(0,\epsilon)}  , \\
		\ket{\Phi^{-}(\epsilon)} \coloneqq &\ket{\psi_E(\pi,\epsilon)} .
	\end{align} 
	\end{subequations}
The corresponding atomic steady states are the pure states, $\hat{\rho}_{\rm ss} = \op{\Phi^\mp(\epsilon)}{\Phi^\mp(\epsilon)}$. The partial transpose matrices $\hat{\rho}_{\rm ss}^{\rm PT}$ have respective eigenvalues 
\beq
\(\frac{1}{2+\epsilon}, \frac{1+\epsilon}{2+\epsilon}, \frac{1}{2}, -\frac{1}{2}  \) ,
\eeq
that guarantee entanglement. The logarithmic negativity of the steady states  is shown in \frf{fig:log_neg} for $\epsilon = 0.001$ (dotted red line).

\begin{figure}[b]
\includegraphics[scale=0.6]{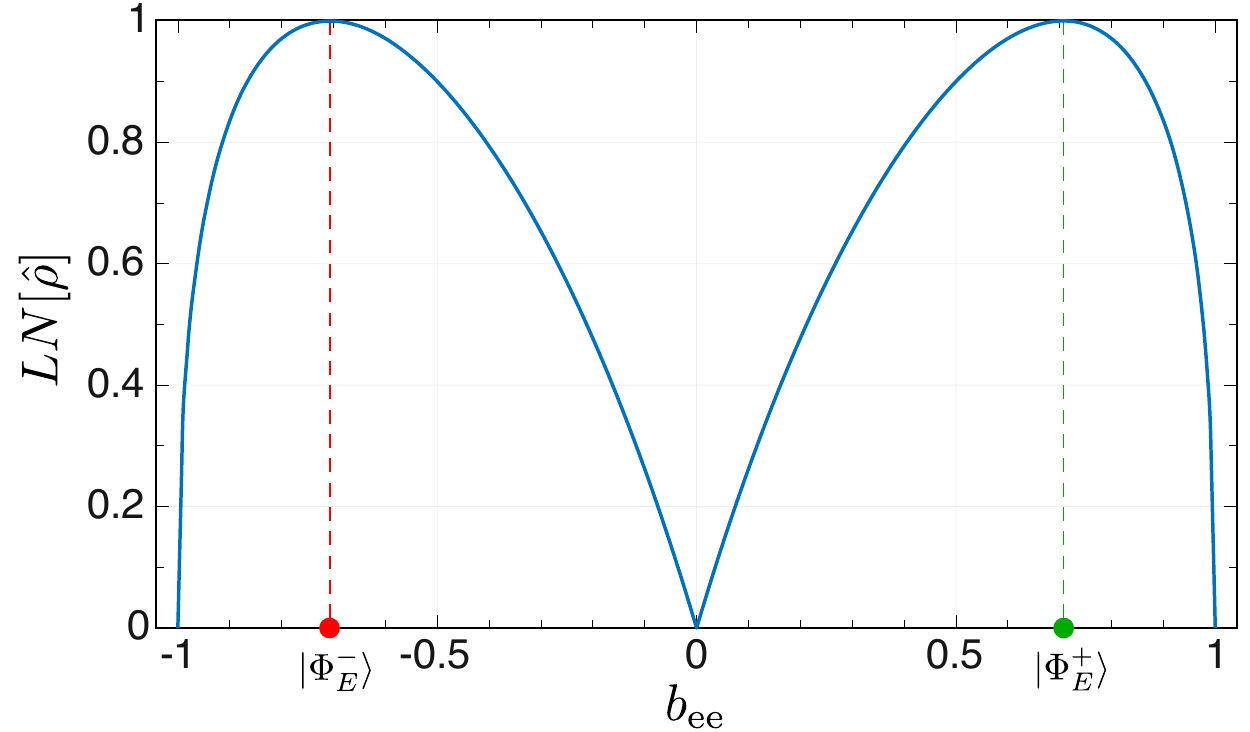}
\caption{\label{fig:log_neg:E} (Color online) Steady-state entanglement of two remote atoms interacting with a two-qubit bath, quantified by the logarithmic negativity $LN(\hat{\rho}_{\rm ss})$. The qubit bath is parameterized by \erf{phi:gen} with $b_{\er \er}$ and $b_{\gc \gc}$ taken to be real. At the exceptional points, where the bath is prepared in a Bell state, $b_{\er \er} = \pm 1/\sqrt{2}$ indicated by red and green dots, the atomic steady-state is not unique. In this case, the steady-state entanglement is a function of the initial atomic state (see \frf{fig:log_neg}).
} 
\end{figure}

This behavior continues for the bath qubits far from Bell states. In Fig. \ref{fig:log_neg:E} we plot the negativity of the atomic steady state for the two-qubit bath state given by \erf{phi:gen}, with $b_{\er \er}$ and $b_{\gc \gc}$ taken to be real.

For $b_{\er\er} = \{0,\pm1\}$ the qubit bath state is separable, and
as expected, the atoms relax to an uncorrelated steady state. 
As the entanglement in the bath qubits increases, so does that of the atomic steady-state. The logarithmic negativity approaches a limiting value of $LN(\hat{\rho}_{\rm ss}) \rightarrow 1$ as the bath approaches a Bell state, $b_{\er\er}= \lim_{\epsilon \rightarrow 0}\pm1/\sqrt{2+\epsilon}$. Indeed, for each value of $b_{\er \er}$, the atomic steady state has the same logarithmic negativity as is present in the two-qubit bath state, indicated full distribution of environmental entanglement to the atoms. As discussed in \srf{Sec:Bellbath} at $b_{\er\er}=\pm1/\sqrt{2}$ (red and green dots), where the environment is \emph{exactly} in the Bell states, the steady-state atoms can exhibit any value of the logarithmic negativity, depending on their initial state.
Recall that a similar situation arose for the case of two remote cavities in \srf{sec:2cav}. We found that when the bath qubits were prepared in the $\ket{\Phi_E^\pm}$ Bell state, this represented the limit of infinite squeezing for the cavities' steady state.

\section{General case: Master equations for $n$-qubit environments} \label{sec:ME}
Our focus thus far has been on the system dynamics and steady-state properties generated by two-qubit environments.
A natural question is: how does the evolution differ when the environmental consists of more than two qubits? In order to address this, we expand the formalism presented in \srf{sec:ME:2q} to the case of $n$-qubit environments. We present the ME for $n$-qubit environments and illustrate some key differences when going beyond two-qubit baths.

Let us now proceed by supposing that all approximations required for the ME derivation in \srf{sec:ME:2q} (Born-Markov and weak-coupling) are applicable here as well. At each interaction time interval the environment is prepared in a general $n$-qubit state $\hat \rho_E$.  It is straightforward to generalize the master equation derived for the two-qubit environments, \erf{ME:gen:2q}, to the $n$-qubit scenario. Following the same procedure~Eq.~\eqref{Utaylor} is substituted into \erf{rhoS:gen} so that the evolution of the joint $n$-system is described by the following ME,
 \bqa \label{nME:gen}
        \dot{\hat \rho}(t) &=& \sum_{\ell=1}^n \(-\, i  \big[ \hat{H}^{\rm eff}_\ell, \hat \rho \big]+  \gd{\ell} \,{\cal D}[\hat{c}_\ell] \hat \rho + \gu{\ell} \,{\cal D}[c\dg_\ell] \hat \rho \) \nonumber \\
        && + \sum_{\ell<m} \Big(\gamma_{\downarrow \downarrow, \ell m} \,{\cal S}[\hat{c}_\ell,\hat{c}_m] + \gamma_{\downarrow \uparrow, \ell m} \,{\cal S}[\hat{c}_\ell,\hat{c}_m\dg]   \nonumber \\
        &&  \quad +\, \gamma_{\downarrow \uparrow, \ell m}^* \,{\cal S}[\hat{c}_\ell\dg,\hat{c}_m] + \gamma_{\downarrow \downarrow, \ell m}^* \,{\cal S}[\hat{c}_\ell\dg,\hat{c}_m\dg] \Big) \hat \rho     ,
\eqa
where
\begin{subequations} \label{gammas:gen:nq}
   \bqa
         \gd{\ell} &=&  \gamma_\ell \mbox{Tr}_E \big[ \bra{\gc_\ell} \hat{\rho}_E \ket{\gc_\ell} \big] , \label{gd:n}  \\
         \gu{\ell} &=& \gamma_\ell \mbox{Tr}_E \big[ \bra{ \er_\ell} \hat{\rho}_E \ket{ \er_\ell} \big] , \label{gu:n}  \\
        \gamma_{\downarrow \downarrow, \ell m} &=&\sqrt{ \gamma_\ell \gamma_m} \mbox{Tr}_E \big[ \bra{ \gc_\ell, \gc_m} \hat{\rho}_E \ket{\er_\ell, \er_m} \big] , \label{gdd:n} \\
         \gamma_{\downarrow \uparrow, \ell m} &=& \sqrt{ \gamma_\ell \gamma_m}  \mbox{Tr}_E \big[ \bra{ \gc_\ell, \er_{m} } \hat{\rho}_E \ket{\er_\ell, \gc_m} \big] , \label{gdu:n} \\
         \hat{H}^{\rm eff}_{\ell} &=&\gamma_\ell \mbox{Tr}_E \big[ \bra{\gc_\ell} \hat{\rho}_E \ket{\er_\ell} \big] \hat{c}_\ell + {\rm H. c.}   \label{Heff:nq}
    \eqa
\end{subequations}
where for example $\ket{\er_\ell, \er_m}$ is the excited bath state for subsystems $\ell$ and $m$. 
The above ME can also be straightforwardly expressed in diagonal, Lindblad form with $2n(n-1)$ jump operators; however, the form above is more amenable when considering mixed-state baths.

\begin{figure}
\includegraphics[scale=0.25]{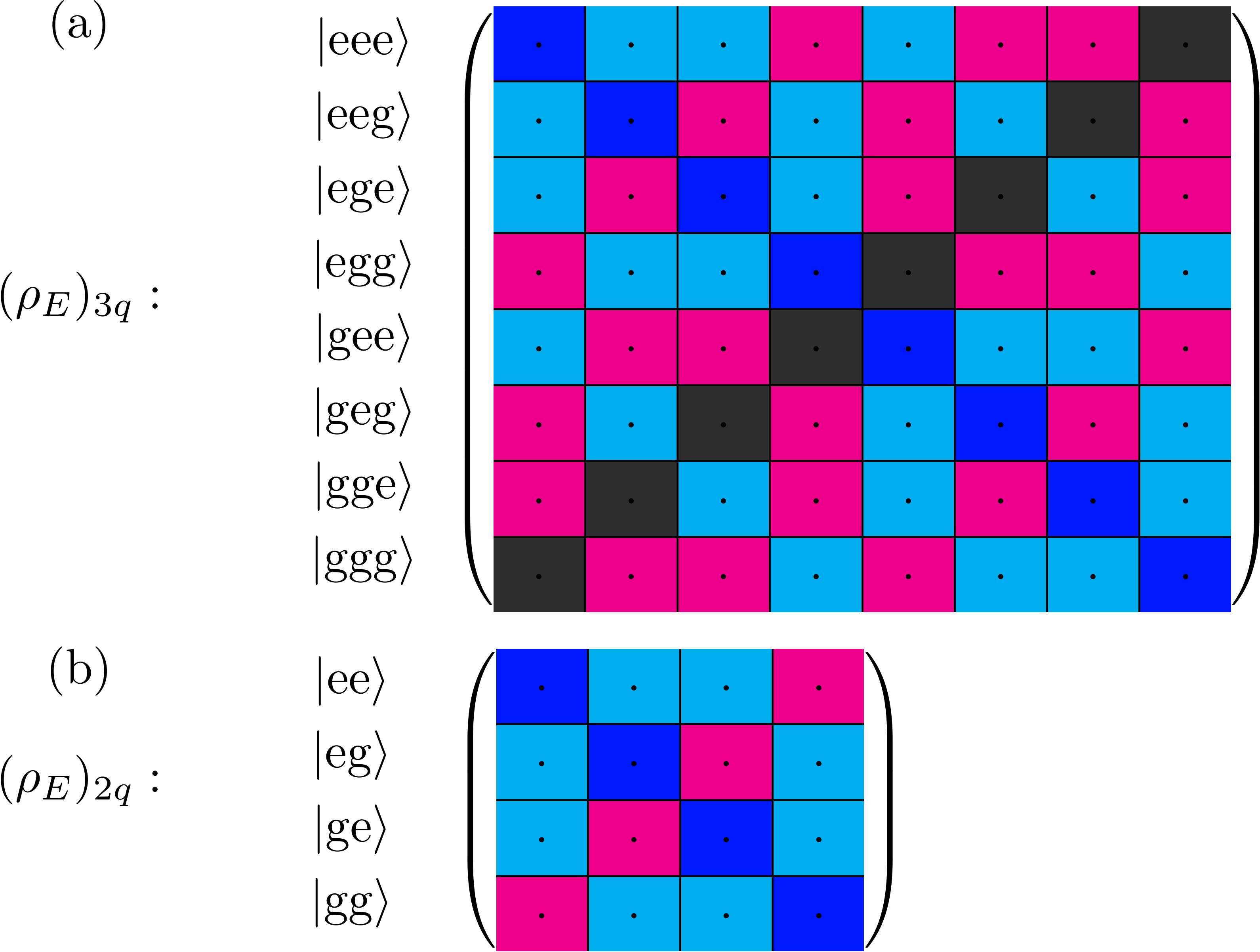}
\caption{\label{fig:rhoE} (Color online). Color-coded representation of the state matrix for three- and two-qubit environments. Matrix elements in the local energy basis contribute to different terms in the multi-qubit ME, Eqs. (\ref{nME:gen}-\ref{gammas:gen:nq}). Matrix elements labeled in cyan (\colorbox{cyan}{$\cdot$}) contribute to the effective Hamiltonians, $\hat{H}^{\rm eff}_\ell$. Those labeled in dark blue (\colorbox{blue}{$\cdot$}) contribute to local dissipators, $\mathcal{D}[\hat{o}] \hat{\rho}$, and those labeled in magenta (\colorbox{magenta}{$\cdot$}) contribute to the two-body superoperators, $\mathcal{S}[\hat{o}_\ell, \hat{o}_m] \hat{\rho} $. Matrix elements labeled in dark grey (\colorbox{darkgray}{$\cdot$}) do not contribute to the ME dynamics. Note the similarities to Fig. 3 in Ref. \cite{DagKur16}.}
\end{figure}

The ME in \erf{nME:gen} has three physical mechanisms (which are indepedent of $n$): (1) local coherent driving of the systems through the effective Hamiltonians, (2) processes involving the gain $(\uparrow)$ or loss $(\downarrow)$ of a system excitation via local dissipators, and (3) two-excitation processes described by superoperators $\mathcal{S}[\hat{o}_\ell, \hat{o}_m] \hat{\rho}$. When the $n$-qubit bath state is expressed in the local energy basis, each matrix element contributes to one of these three processes through the coefficients in \erf{gammas:gen:nq} or does not drive dynamics at all. Indicated in \frf{fig:rhoE} are the contribution of the environmental state matrix elements to each of these processes (in cyan, dark blue, and magenta, respectively), for two- and three-qubit environments. 
Consider first the $n=3$ bath. An important observation is the absence of antidiagonal matrix elements, indicated by dark grey cells in \frf{fig:rhoE}(a), in any coefficients in \erf{gammas:gen:nq}. The fact that the antidiagonal matrix elements do not affect the ME dynamics is indicated by the dark grey cells in \frf{fig:rhoE}. Surprisingly, this implies that preparing the environmental qubits in certain maximally entangled states such as the Greenberger--Horne--Zeilinger (GHZ) state \cite{GHZ07} in the local energy basis is not useful for generating correlations between the subsystems for $n>2$. Additionally, if the quantum systems are atoms prepared in a GHZ state, their nonclassical quantum correlations eventually decohere, as expected for three subsystems subject to local dissipation such as dephasing or depolarizing channels \cite{Wei10}. This is in marked contrast to two-qubit baths where the antidiagonal components of $\hat{\rho}_E$ that appear in Eqs. (\ref{gdd}-\ref{gdu}), indicated by the magenta cells in \frf{fig:rhoE}(b), are the key ingredient for entangling the systems. This highlights the significance of the two-qubit $X$-state environments (described by a state matrix in which only diagonal and antidiagonal entries are nonzero) \cite{Wei10, RafEbe12}. 
Da\v{g} \emph{et al.} \cite{DagKur16} encountered a similar result---they found that antidiagonal coherences in $n=3$ qubit baths do not contribute to squeezing or displacement of a single cavity mode.

This behavior also extends to $n$-qubit baths. The anti-diagonal matrix elements of $\hat{\rho}_E$ have the form $\ket{x}\!\bra{\bar{x}}$ where $x$ is a string of $\er$ and $\gc$ labels and $\bar{x}$ is the complement string with $\er$ and $\gc$ swapped. Then for $n\ge 3$ and all valid $\ell$ and $m$ in \erf{gammas:gen:nq}, either the inner product is zero or the trace over the remaining systems is zero. The coherences in the $n$-qubit $X$-state bath that arise from this type of more than two-body entanglement among the qubits do not contribute to the dynamics in the ME.  As a result, such maximally entangled states do not play any role in entangling subsystems.


Similar to \srf{prod:bath} one might ask whether or not the presence of cross terms in the ME guarantees the generation of entanglement among the subsystems? The short answer is \emph{no} which is supported by the following counter example. Assume that the environment is in a pure product state $\ket{\psi_E} = \ket{\psi_\ell}^{\otimes n}$ where $\ket{\psi_\ell} = \alpha_\ell \ket{\gc} + \beta_\ell \ket{\er}$ with $|\alpha_\ell|^2 + |\beta_\ell|^2=1$, and the system is also initially in a product state $\hat\rho = \sum_{\ell=1}^n \hat\rho_\ell^{\otimes n}$. For this setting, the dynamics of the joint system is governed by the following ME
 \begin{align} \label{nME:prod}
        \dot{\hat \rho}(t) &= -i \sum_{\ell=1} \hat\rho_1 \otimes \cdots \otimes \hat\rho_{\ell-1}\otimes \Big[ H^{\rm eff}_\ell, \hat\rho_\ell \Big] \otimes \hat\rho_{\ell+1} \otimes  \nonumber \\
        & \cdots \otimes \hat\rho_n + \sum_{\ell=1} \gamma_\ell \Big( |\alpha_\ell|^2 \mathcal{D}[\hat{c}_\ell] +|\beta_\ell|^2 \mathcal{D}[\hat{c}_\ell\dg] \Big) \hat{\rho}    \\
        & -\sum_{\ell<m} \, \sqrt{\gamma_\ell \gamma_m} \, \hat\rho_1 \otimes \cdots \otimes \hat\rho_{\ell-1} \otimes \Big[ H^{\rm eff}_\ell, \hat\rho_\ell \Big] \otimes \nn \\
        & \hat\rho_{\ell+1} \otimes \cdots \otimes \hat\rho_{m-1} \otimes \Big[ H^{\rm eff}_m, \hat\rho_m \Big] \otimes \hat\rho_{m+1} \otimes \cdots \otimes \hat\rho_n, \nn
\end{align}
with effective Hamiltonians defined in \erf{ham:eff:prod}. The same line of discussion presented in \srf{prod:bath} holds here. That is, the first two summations describe the independent evolution of each subsystem, and the last one describes a classically correlated driving of the all possible two-body configurations. The latter does not lead to entanglement creation so that the system keeps its product state structure.

This analysis suggests a way that could be used for steady-state entanglement across all subsystems, which could be verified by calculating an $n$-partite entanglement witness \cite{BarBla13, SpeVog13, PezSme16, GirYad17, GhiMac18}. Since the ME, given in \erf{nME:gen}, has only two-body cross terms, generating entanglement among the subsystems might be achieved by engineering a specific structure in $\hat\rho_E$. It seems the presence of pairwise entanglement in a particular form in the bath qubits is sufficient to make sure that the coefficients in \erfa{gdd:n}{gdu:n} are nonzero. It remains an open question whether such an environment with pair-wise entanglement would enable entangling dynamics in the ME, or perhaps more generally a sequence of entangled environments operating for consecutive periods of time. In addition, the cost of engineering such an environment would be an important factor.


Although the absence of coupling from anti-diagonal coherences in $\hat{\rho}_E$ may appear puzzling, some of the observed structure will be due to the weak-coupling approximation that underpins the master equation derivation. Recall that the joint state of the systems is updated by a dynamical map, \erf{rhoS:gen}, that results from tracing out the bath after unitary evolution over a small interaction time $\Delta t$. In the weak coupling regime the expansion of the unitary operator, \erf{Utaylor}, is truncated at the second-order term in $\Delta t$. This limits the influence of bath correlations in the ME to two-body terms. If one were to keep terms up to $\Delta t^3$ or beyond, these coherences would play a part in the dynamical map. This could arise when the coupling time between each successive qubit and its corresponding subsystem is large enough that weak coupling criterion, $\lambda_i \Delta t \ll 1$, is not entirely valid and requires perturbative corrections.


\section{Conclusion} \label{sec:con}
Within the repeated quantum interaction formalism we have derived master equations for open quantum systems evolving under irreversible entangled quantum channels. 
The environment is composed of a chain of identical entangled two-level systems which sequentially interact weakly with their corresponding subsystems and are then discarded. 
In the limit of a continuous stream of environmental qubits, the joint system evolves according to a Markovian ME with local effective Hamiltonians and nonlocal dissipative processes with jump operators that are combinations of creation and destruction operators {\em across} the subsystems. This description applies generally to a qubit bath in a mixed state,  
for which we provide an alternate, nondiagonal form of the ME that can be easier to work with.

A pedagogical study of a two-qubit bath coupled to a pair of two-level subsystems led to several conclusions. 
First, we find that the presence of antidiagonal coherences in the bath is a necessary but insufficient condition for steady-state entanglement of the subsystems, when the bath state is expressed in a basis of local eigenstates.
For instance, the bath can be prepared in a product state with each qubit in a superposition of $\ket{\gc}$ and $\ket{\er}$. While coherences exist, the stationary state of the system is not entangled. 
Second, maximally entangled bath states do not give rise to unique steady states while even slight deviations from these ``exceptional" bath states do.

For the general case of entangled $n$-qubit baths, the ME contains at most two-body terms in the jump operators. 
A surprising consequence is that particular maximally entangled $n$-qubit baths (for $n > 2$) do not affect subsystem entanglement.  
That is, when expressed in the local energy basis antidiagonal coherences in the bath state do not couple to the system. 
An implication is that $X$-state baths drive the same dynamics as diagonal state baths such as a thermal bath. 
This is in contrast to the two-qubit environments where the existence of antidiagonal coherences are essential to the generation of entanglement between the systems. 

This work opens several avenues for future research. Extending the methods of Gross \emph{et al.} \cite{GroCom17}, the formalism presented here offers a way to model multimode Gaussian bosonic baths such as two-mode squeezed electromagnetic environments. Thermal and other mixed-state baths can be directly modeled by tracing over a part of a pure multi-qubit entangled bath. 
Moreover, it is possible to proceed beyond Gaussian baths by perturbatively extending the weak coupling limit with
the inclusion of higher order terms, ${O}[(\lambda \Delta t)^k]$, in the interaction Hamiltonian.

Environment-assisted entangling protocols based on engineered qubit environments that include {\em only} pairwise entanglement across all qubits could be used to create useful multipartite entanglement among all systems (generating cluster states, for example). This might be useful specially when entangling multipartite system is practically challenging. 
Several extensions to the bath itself could be studied. The bath qubits could be replaced by $d$-dimensional quantum systems, or qudits, yielding a richer structure to the environment, and simultaneous spatial entanglement between qubits across channels and entanglement-in-time between progressive qubits in a single channel such that the evolution is inherently nonlocal and non-Markovian \cite{Baragiola:2012aa, LorPal15, DabChr17}. 

\section*{Acknowledgments}
B.Q.B. thanks Nicolas Menicucci and Rafael Alexander for valuable discussions. 
This project was supported in part by the Australian Research Council (ARC) Centres of Excellence for Engineered Quantum Systems (CE110001013, CE170100009). B.Q.B. also acknowledges support from the ARC Centre of Excellence for Quantum Computation and Communication Technology (Project No. CE170100012).

	\appendix 
	\begin{widetext}
	\numberwithin{equation}{section}
	\section{Master equation derivation for a two-qubit environment} \label{appnA1}
	Inserting the unitary interaction, $\hat{U}_I^{(l)}$, from \erf{Utaylor} into the dynamical map in \erf{rhoS:gen} and keeping terms to second order in $\Delta t$, the following expression is obtained \cite{DagKur16, LorPal15, LiSha18},
	\bqa
	\hat \rho{(t_{l+1})} &=& {\rm Tr}_E \bigg( \hat \rho{(t_l)} \otimes \hat{\rho}_E - i \Delta t\, \big[ \hat{H}^{(l)}_I,  \hat \rho(t_l) \otimes \hat{\rho}_E \big]+ {\Delta t}^2 \, \hat{H}^{(l)}_I\, \hat \rho(t_l) \otimes \hat{\rho}_E \hat{H}^{(l)}_I - \frac{{\Delta t}^2}{2} \big\{ (\hat{H}^{(l)}_I)^2 ,\hat \rho(t_l) \otimes \hat{\rho}_E \big\}_+ \bigg),
	\eqa
where $\{ \hat A, \hat B \}$ denotes an anticommutator. We will explicitly take the environmental trace for each term in the expansion with respect to the environment, $\hat{\rho}_E = \op{\psi_E}{\psi_E}$, where $\ket{\psi_E}$ is the two-qubit state given in \erf{rhoE:pure}. The first term is simply just the system state $\hat{\rho}(t_l)$.  The commutator term proportional to $\Delta t$ becomes
	\beq
	{\rm Tr}_E \( \tr{\hat{H}^{(l)}_I,  \hat{\rho}(t_l) \otimes \hat{\rho}_E} \) =  \left[ \lambda_1 \hat{c}_1 \(b_{\gc\gc} b_{\er \gc}^\ast  + b_{\gc \er} b_{\er\er}^\ast\) + \lambda_2 \hat{c}_2 \(b_{\er\gc} b_{\er\er}^\ast+ b_{\gc\gc} b_{\gc \er}^\ast\) + {\rm H. c.}, \,  \hat{\rho}(t_l) \right].
	\eeq
Now we turn our attention to the second-order terms, proportional to $\Delta t^2$. The first term in the second line becomes
	\bqa
	{\rm Tr}_E \Big(  \hat{H}^{(l)}_I\, \hat{\rho}(t_l) \otimes  \hat{\rho}_E  \hat{H}^{(l)}_I \Big) && =  \lambda_1^2  \left\{\({\abs{b_{\gc\er}}^2 + \abs{b_{\gc\gc}}^2}\) \hat{c}_1 \hat{\rho}(t_l) \hat{c}_1\dg + \(\abs{b_{\er\er}}^2 + \abs{b_{\er\gc}}^2\)  \hat{c}_1\dg \hat{\rho}(t_l) \hat{c}_1  \right\}  \nn \\
	&& +\lambda_2^2  \left\{\({\abs{b_{\er\gc}}^2 + \abs{b_{\gc\gc}}^2}\) \hat{c}_2 \hat{\rho}(t_l) \hat{c}_2\dg + \(\abs{b_{\er\er}}^2 + \abs{b_{\gc\er}}^2\)  \hat{c}_2\dg \hat{\rho}(t_l) \hat{c}_2  \right\} \\
	&& +\lambda_1 \lambda_2  \left\{ b_{\gc\gc} b_{\er\er}^\ast \, \hat{c}_1  \hat{\rho}(t_l) \hat{c}_2 + b_{\gc\er} b_{\er\gc}^\ast \, \hat{c}_1  \hat{\rho}(t_l) \hat{c}_2\dg + b_{\er\er} b_{\gc\gc}^\ast \, \hat{c}_1\dg \hat{\rho}(t_l) \hat{c}_2\dg + b_{\er\gc} b_{\gc\er}^\ast \, \hat{c}_1\dg \hat{\rho}(t_l) \hat{c}_2 + {\rm H. c.} \right\}. \nonumber
	\eqa
And the remaining anti-commutator term becomes
	\bqa
	 {\rm Tr}_E \Big[ \big\{ (\hat{H}^{(l)}_I)^2,  && \hat{\rho}(t_l)  \otimes    \hat{\rho}_E \big\}_+ \Big] = \lambda_1^2  \, \left\{\(\abs{b_{\er\er}}^2 + \abs{b_{\er\gc}}^2\) \big\{ \hat{\rho}(t_l) , \hat{c}_1 \hat{c}_1\dg \big\}_+ + \(\abs{b_{\gc\er}}^2 + \abs{b_{\gc\gc}}^2\) \big\{ \hat{\rho}(t_l), \hat{c}_1\dg \hat{c}_1 \big\}_+ \right\} \nn  \\
	  &&\hspace{10pt } + \lambda_2^2 \, \left\{\(\abs{b_{\er\er}}^2 + \abs{b_{\gc\er}}^2\) \big\{ \hat{\rho}(t_l), \hat{c}_2 \hat{c}_2\dg \big\}_+ + \(\abs{b_{\er\gc}}^2 + \abs{b_{\gc\gc}}^2\) \big\{ \hat{\rho}(t_l), \hat{c}_2\dg \hat{c}_2 \big\}_+ \right\}  \\
	 && \hspace{10pt } + 2 \lambda_1 \lambda_2  \, \left\{ b_{\gc\gc} b_{\er\er}^\ast \big\{ \hat{\rho}(t_l), \hat{c}_1 \hat{c}_2 \big\}_+ + b_{\gc\er} b_{\er\gc}^\ast \big\{ \hat{\rho}(t_l) ,\hat{c}_1 \hat{c}_2\dg \big\}_+ + \,b_{\er\er} b_{\gc\gc}^\ast \big\{ \hat{\rho}(t_l), \hat{c}_1\dg \hat{c}_2\dg \big\}_+ + b_{\er\gc} b_{\gc\er}^\ast \big\{ \hat{\rho}(t_l), \hat{c}_1\dg \hat{c}_2 \big\}_+ \right\}. \nonumber 
	\eqa

 Now collecting these terms and assuming that the time interval $\Delta t = t_{l+1} - t_l$ is small enough to make the following approximation
	 \[\dot{ \hat{\rho} }(t) \equiv \lim_{\Delta t \rightarrow 0} \frac{\hat \rho{(t_{l+1})} - \hat \rho{(t_{l})}}{t_{l+1} - t_l}, \] we can construct the ME given by \erf{ME:gen}.

\end{widetext}

\section{Gaussian evolution} \label{appendix:Gaussian}

Here we briefly review the description of multimode Gaussian bosonic states and their open systems evolution. Consider a system composed of $N$ bosonic modes. We work in the basis of Hermitian position  and momentum operators for each mode, $\hat{q} = \frac{1}{\sqrt{2}}(\hat{a} + \hat{a}\dg)$ and $\hat{p} = \frac{1}{i\sqrt{2}}(\hat{a} - \hat{a}\dg)$, and the canonical commutation relations are $[\hat{q}_\ell, \hat{p}_m] = \delta_{\ell m}$. Noting the commutation relation for vectors of operators,
$[\opvec{r}, \opvec{s}^{\top}] = \opvec{r} \opvec{s}^{\top} - (\opvec{s} \opvec{r}^{\top})^{\top} $ where ${}^\top$ denotes matrix transpose, we define a column vector of stacked position and momentum operators, $\opvec{x} \coloneqq (\opvec{q} \;\opvec{p})^\top$. 
Then, the canonical commutation relations can be succinctly stated as
	\begin{align} \label{eq::commutationsqp}
		\sq{\opvec{x}, \opvec{x}^{\top}} = i \mat{\Omega},
	\end{align}
where the matrix $\mat{\Omega}$ is called the symplectic form and has the following representation in the $qp$-basis:
	\begin{align} \label{eq::symplecticform}
		\mat{ \Omega} = 
		\left(
		\begin{array}{cc}
			\mat{0} & \mat{I} \\
			-\mat{I} & \mat{0} 
		\end{array}
		\right),
	\end{align}
with $\mat{I}$ being the $N\times N$ identity matrix. 

A Gaussian quantum state of $N$ bosonic modes is entirely described by a vector of means $\vec{x} \coloneqq \expt{\opvec{x}}$ and a symmetrized covariance matrix $\mat{\Sigma}$ 
with elements $\Sigma_{ij} = \frac{1}{2} \langle \hat{x}_i \hat{x}_j + \hat{x}_j \hat{x}_i \rangle$. The covariance matrix for the vacuum state is $\mat \Sigma_0 = \frac{1}{2} \mat{I}$. 
 The squeezing operator in \erf{eq:squeezeop} was defined with respect to its Schr\"{o}dinger-picture action; to find the covariance matrix we need its Heisenberg-picture action. Using $\hat{S}\dg(r, \vartheta) = \hat{S}(r,\vartheta+\pi)$,
the covariance matrix for a two-mode squeezed state can be found from its associated symplectic matrix \cite{Menicucci:2011aa},
	\begin{align} \label{eq::covMat2MS}
		\mat{ \Sigma }(r)  
		= & \frac{1}{2} \left(
		\begin{array}{cccc}
			\cosh (2r) & -\sinh (2r) & 0 & 0 \\
			-\sinh (2r) & \cosh (2r) &0 & 0 \\ 
			0 & 0 & \cosh (2r) & \sinh (2r) \\ 
			0 & 0 & \sinh (2r) & \cosh (2r) 
		\end{array}
		\right), 
	\end{align}
where we set $\vartheta = 0$ for simplicity.

Gaussianity of a quantum state is preserved under evolution generated by a Hamiltonians quadratic in the mode operators and jump operators that are linear in the mode operators---the master equation in \erf{ME:gen} is one such example. Given a Lindblad master equation with $M$ jump operators that governs the evolution of a multimode bosonic state,
	\begin{align} \label{eq:vacME}
		 \dot{\hat{ \rho }}
		 = - i [\hat{H}, \hat{\rho}] + \sum_{m=1}^M \mathcal{D}[\hat{L}_m] \hat{\rho},
	\end{align}
we may translate this into an evolution for the means and the covariance matrix without loss of information if Gaussianity is preserved. The Gaussian-preserving conditions require that the Hamiltonian can be written as
	\begin{align}
		\hat{H} = \smallfrac{1}{2} \opvec{x}^{\top} \mat{G} \opvec{x},
	\end{align}
expressed in terms of the symmetric, real matrix $\mat{G} \in \mathbb{R}^{2N \times 2N}$, and each jump operator has the form 
	\begin{align} \label{Eq::JumpOps}
		\hat L_m = \sqrt{\gamma_m}  \sum_{\ell = 1}^N \big( Q_{m\ell} \hat q_\ell + P_{m\ell} \hat p_\ell \big),
	\end{align}
where $\gamma_m$ is the associated dissipation rate. Collecting the $Q_{m\ell}$ and $P_{m\ell}$ coefficients into the matrices $\mat{Q}$ and $\mat{P}$, the vector of $M$ jump operators is 
	\begin{align} \label{Eq::Lmatrix}
		\opmat{L} = \mat{C} \opvec{x},
	\end{align}
with $\mat{C} \coloneqq  ( \mat{Q} \, \,  \mat{P} )  \in \mathbb{C}^{M \times 2N}$.
The means and covariance matrix 
obey the following equations of motion \cite{Koga:2012aa}:
	\begin{align}
		\dot{\vec{x}} =& \mat{A} \vec{x}, \\
		\dot{\mat{\Sigma}} =& \mat{A} \mat{\Sigma} + \mat{\Sigma} \mat{A}^{\top} + \mat{B} , \label{Eq::CovEOM}
	\end{align}
with matrices
	\begin{align}	
		\mat{A} &= \mat{\Omega} \big( \mat{G} + \mathfrak{Im} [\mat{C}^{\rm H} \mat{C}] \big), \\
		\mat{B} &= \mat{\Omega} \, \mathfrak{Re} [\mat{C}^{\rm H} \mat{C}] \, \mat{\Omega}^{\top} , \label{eq:covmatEOM}
	\end{align}	
where ${}^{\rm H}$ indicates matrix conjugate transpose (to distinguish it from the Hermitian adjoint $\dg$ of an operator). 
For the two-mode ME in \srf{sec:2cav} $\mat{G} = \mat{0}$ and the jump operators are given by \erf{sq:op}. 
Setting $\vartheta = 0$ the matrices in the covariance matrix evolution, \erf{Eq::CovEOM}, are $\mat{ A }  = \ \frac{\Gamma}{2} \mat{I}$ and $\mat{B} =  \Gamma \mat{\Sigma}(r).$

The Uhlmann-Jozsa fidelity 
	\begin{align}
		\mathcal{F}(\hat{\rho}, \hat{\varrho}) = \left[ \text{tr}\left( \sqrt{\sqrt{\hat\rho} \hat \varrho \sqrt{\hat \rho} }\right) \right]^2
	\end{align}
is a measure of the closeness of the quantum states $\hat{\rho}$ and $\hat{\varrho}$. When both states are Gaussian at least one is pure, the fidelity can be calculated directly from their respective covariances matrices, $\mat{\Sigma}_\rho$ and $\mat{\Sigma}_\varrho$, 
	\begin{align}
		\mathcal{F}(\hat{\rho}, \hat{\varrho}) = \left[ \det \left( \mat{\Sigma}_\rho + \mat{\Sigma}_\varrho \right) \right]^{-1/2}
	\end{align}
where $\det(\mat{A})$ is the determinant of the matrix $\mat{A}$  \cite{Spedalieri:2013aa}. In \srf{sec:2cav} the two-mode squeezed state is pure with a covariance matrix given by \erf{eq::covMat2MS}.

\twocolumngrid
\bibliography{references}

\end{document}